\documentclass[aps,prl,twocolumn,superscriptaddress]{revtex4-2}

\usepackage{mathrsfs,amssymb,graphics,subfigure,threeparttable}
\usepackage{grap hicx}
\usepackage{amssymb}
\usepackage{enumerate}
\usepackage{amsmath}
\usepackage{fancyhdr}
\usepackage{bm}
\usepackage[squaren]{SIunits}
\usepackage{hyperref}

\usepackage{xcolor}
\usepackage{soul}
\begin{document}

%\preprint{}

\title{Bidirectional cascaded superfluorescent lasing in air enabled by resonant third harmonic photon exchange from nitrogen to argon}


\author{Zan Nie}
\email[]{znie@hust.edu.cn}
\affiliation{Department of Electrical and Computer Engineering, University of California Los Angeles, Los Angeles, California 90095, USA}
\affiliation{Wuhan National Laboratory for Optoelectronics, Huazhong University of Science and Technology, Wuhan 430074, China}

\author{Noa Nambu}
\author{Kenneth A. Marsh}
\author{Daniel Matteo}
\author{C. Kumar Patel}
\author{Chaojie Zhang}
\author{Yipeng Wu}
\affiliation{Department of Electrical and Computer Engineering, University of California Los Angeles, Los Angeles, California 90095, USA}

\author{Stefanos Carlstr\"om}
\author{Felipe Morales}
\author{Serguei Patchkovskii}
\author{Olga Smirnova}
\author{Misha Ivanov}
\email[]{Mikhail.Ivanov@mbi-berlin.de}
\affiliation{Max Born Institute, Max-Born-Str. 2A, D-12489 Berlin, Germany}

\author{Chan Joshi}
\email[]{cjoshi@ucla.edu}
\affiliation{Department of Electrical and Computer Engineering, University of California Los Angeles, Los Angeles, California 90095, USA}

\date{\today}

\begin{abstract}
Cavity-free lasing in atmospheric air has stimulated intense research towards fundamental understanding of underlying physical mechanisms. In this Letter, we identify a new mechanism -- third harmonic photon mediated resonant energy transfer pathway leading to population inversion in argon via initial three-photon excitation of nitrogen molecules irradiated by intense 261 nm pulses -- that enables bidirectional two-color cascaded lasing in atmospheric air. By making pump-probe measurements, we conclusively show that such cascaded lasing results from superfluorescence (SF) rather than amplified spontaneous emission (ASE). Such cascaded lasing with the capability of producing bidirectional multicolor coherent pulses opens additional possibilities for remote sensing applications.

\end{abstract}


\maketitle
“Air lasing refers to the remote optical pumping of the constituents of ambient air that results in a directional laserlike emission from the pumped region.”\cite{laurain_low-threshold_2014} This remarkable phenomenon of cavity-free air lasing discovered about a decade ago \cite{dogariu_high-gain_2011} has opened unique opportunities in remote sensing\cite{hemmer_standoff_2011}, LIDAR, and coherent Raman spectroscopy\cite{malevich_stimulated_2015,zhang_high-sensitivity_2022,fu_air-laser-based_2022, zhang_electronicresonanceenhanced_2023}, especially if robust backward lasing can be achieved. In all schemes demonstrated so far\cite{dogariu_high-gain_2011, yao_high-brightness_2011, zhang_rotational_2013, laurain_low-threshold_2014, liu_recollision-induced_2015, xu_sub-10-fs_2015, dogariu_three-photon_2016, yao_population_2016, polynkin_air_2018, yuan_recent_2019, ando_rotational_2019, danylo_formation_2019, li_significant_2019,li_giant_2020,richter_rotational_2020, zhang_coherent_2021,kleine_electronic_2022,zhuang_optical_2022}, such backward lasing in atmospheric air has been achieved only by resonant excitation of atomic species, such as two-photon excitation of atomic oxygen or nitrogen\cite{dogariu_high-gain_2011, laurain_low-threshold_2014,dogariu_remote_2015} or three-photon excitation of atomic argon (Ar) \cite{dogariu_backwards_2018} with ultraviolet (UV) lasers. To achieve lasing in atomic oxygen and nitrogen, the molecular oxygen (O$_2$) and nitrogen (N$_2$) present in air have to be dissociated first, which induces significant fluctuations of lasing amplitude due to the highly nonlinear intensity dependence of the dissociation process. To avoid such complexity, one needs to rely on the atomic species already present in ambient air, making Ar, the most abundant atomic species in the atmosphere, the best candidate\cite{dogariu_three-photon_2016}. However, Ar concentration in ambient air is only $\sim$1\%, making generation of significant gain required for robust backward lasing in atmospheric air extremely challenging. Here, we solve this problem by using multiphoton excitation of N$_2$, the most abundant atmospheric species, which significantly enhances the excitation efficiency. The created excitation is then efficiently transferred to Ar via third harmoinc photon exchange between the two species, resulting in robust cascaded two-color lasing in atmospheric air. The photon-mediated energy transfer from N$_2$ to Ar occurs on a much faster time scale than the collisional transfer from Ar to N$_2$ in a process known as FLEET\cite{grib_resonance-enhanced_2021,michael_femtosecond_2011}. 

The identification of a new physical mechanism responsible for air lasing in the forward and backward directions is central to understanding the key physics involved. Previous studies attributed lasing in pure Ar to amplified spontaneous emission (ASE)\cite{dogariu_three-photon_2016,zhuang_optical_2022}. However, in experiments carried out under similar conditions, our pump-probe measurement results show that lasing either in pure Ar or in Ar contained in atmospheric air is due to superfluorescence (SF). This is a manifestation of what Dicke called superradiance, i.e., cooperative spontaneous emission\cite{dicke_coherence_1954}. Our measurements show that while the lasing intensity is proportional to the square of the density of excited atoms, both the delay time ($\tau_D$) and duration ($\tau_R$) of the radiated pulses are approximately inversely proportional to the density of excited atom density as expected for SF emission. The condition for pure SF to occur is $\tau_D<T_2$\cite{maki_influence_1989}, where $T_2$ is the dephasing time, which acts to destroy the coherence. Experimentally, in pure Ar, the condition of $\tau_D<T_2$ is easily met. However, in atmospheric air, collisions in atmospheric condition could greatly decrease $T_2$ leading to the quenching of the SF. The new excitation mechanism we discovered, the third harmonic photon mediated resonant energy transfer from N$_2$ to Ar, greatly reduces the characteristic delay time $\tau_D$, allowing bidirectional cascaded SF to occur under atmospheric conditions.

\begin{figure}[tp]
\includegraphics[width=0.5\textwidth]{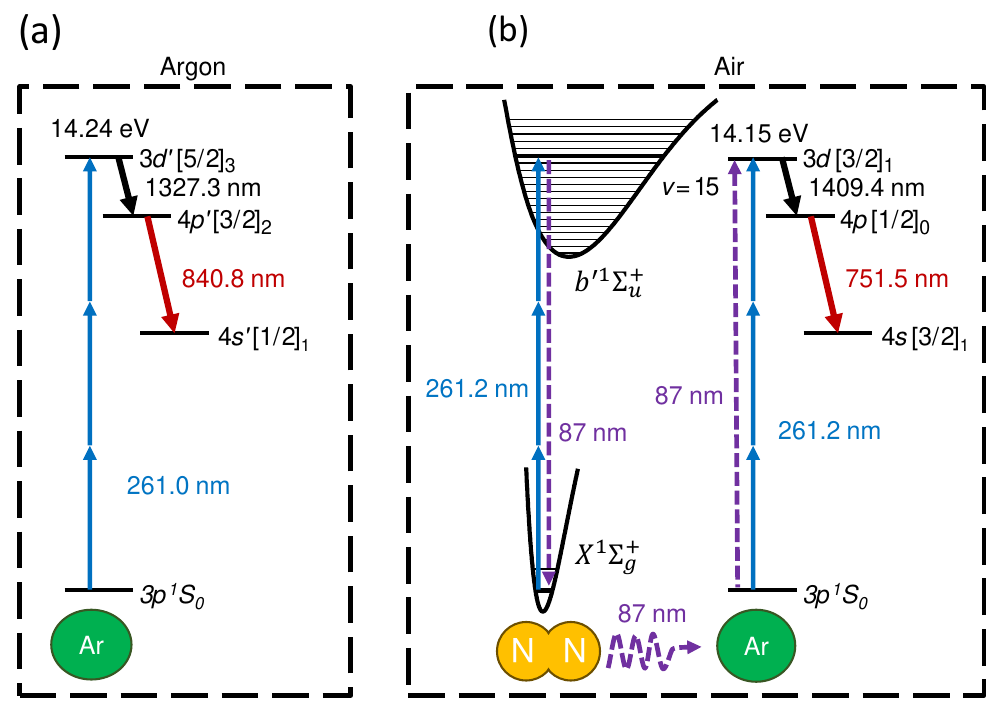}
\caption{\label{fig1} 
Relevant transitions of three-photon excitation and cascade lasing in pure Ar and atmospheric air. (a) In pure Ar, the Ar atoms are excited by the 261 nm pump pulse to the 3d$^\prime$[5/2]$_3$ state (14.24 eV) and emit 1st lasing at 1327.3 nm and 2nd lasing at 840.8 nm. (b) In atmospheric air, the N$_2$ molecules are excited to the b$^{\prime 1}\Sigma_u^+$ (v=15) state (14.1551 eV) by the 261.2 nm pump pulse, and emit third-harmonic photons at 87 nm, which is reabsorbed by the surrounding Ar atoms so that the Ar atoms are excited to the 3d[3/2]$_1$ state (14.1525 eV) and then emit 1st lasing at 1409.4 nm and 2nd lasing at 751.5 nm. Note that the center wavelengths of the pump are slightly different: 261.0 nm (Ar) v.s. 261.2 nm (air) to optimize the pumping of the three-photon resonance.
}
\end{figure}

In pure Ar, excitation to the 3d$^\prime$[5/2]$_3$ state occurs via three-photon absorption of the 261 nm radiation as shown in Fig. \ref{fig1}(a). Superfluorescent lasing occurs on cascaded 1327.3 nm (1st lasing) and 840.8 nm (2nd lasing) transitions. In air, or in Ar and N$_2$ mixtures, increasing gas pressure increases collisional relaxation of the 3d$^\prime$[5/2]$_3$ state, quenching the cascaded lasing from this state. However, we find that efficient population of the 3d[3/2]$_1$ state of Ar atoms can be enabled by initial resonant three-photon excitation of N$_2$ molecules to the b$^{\prime 1}\Sigma_u^+$ (v=15) state, followed by ultrafast energy transfer to the 3d[3/2]$_1$ state of Ar via emission of third harmonic photons and reabsorption by the surrounding Ar atoms as seen from Fig. \ref{fig1}(b). This transfer is reminiscent of the ultrafast interatomic Coulombic decay\cite{cederbaum_giant_1997, santra_interatomic_2000, averbukh_mechanism_2004, jahnke_experimental_2004}, where highly efficient resonant transfer of excitation is also mediated by (virtual) photon exchange and occurs on femtosecond time scale. Importantly, resonantly enhanced emission of third harmonic photons in N$_2$ increases with increasing its pressure. Hence, the energy transfer rate to Ar also grows with pressure, counteracting collisional losses. Subsequently, superfluorescent lasing occurs on cascaded 1409.4 nm (1st lasing) and 751.5 nm (2nd lasing) transitions. 

The experimental setup is described in Supplemental Material. First, we focus on the forward cascade lasing. Resonant three-photon absorption can populate several sublevels of 3d, 3d$^\prime$, and even 5s$^\prime$ orbitals because of the $\sim$9 THz bandwidth of the 80 fs (FWHM) UV pump pulses. We scan the UV pump wavelength to optimize the cascade lasing signals (see Supplementary Fig. 2). On tuning the pump wavelength to 261.0 nm, we have observed lasing in pure Ar at 1327.3 nm and 840.8 nm. This two-step cascade lasing, shown in Fig. \ref{fig1}(a), proceeds via a common energy level 4p$^\prime$[3/2]$_2$ with no other intermediate level(s). When replacing pure Ar with laboratory air at atmospheric pressure (typical humidity 50\%, about 7 Torr partial pressure of Ar), the 1st and the 2nd lasing wavelengths switched to 1409.4 nm and 751.5 nm respectively, as shown in Fig. \ref{fig1}(b). 

\begin{figure}[tp]
\includegraphics[width=0.5\textwidth]{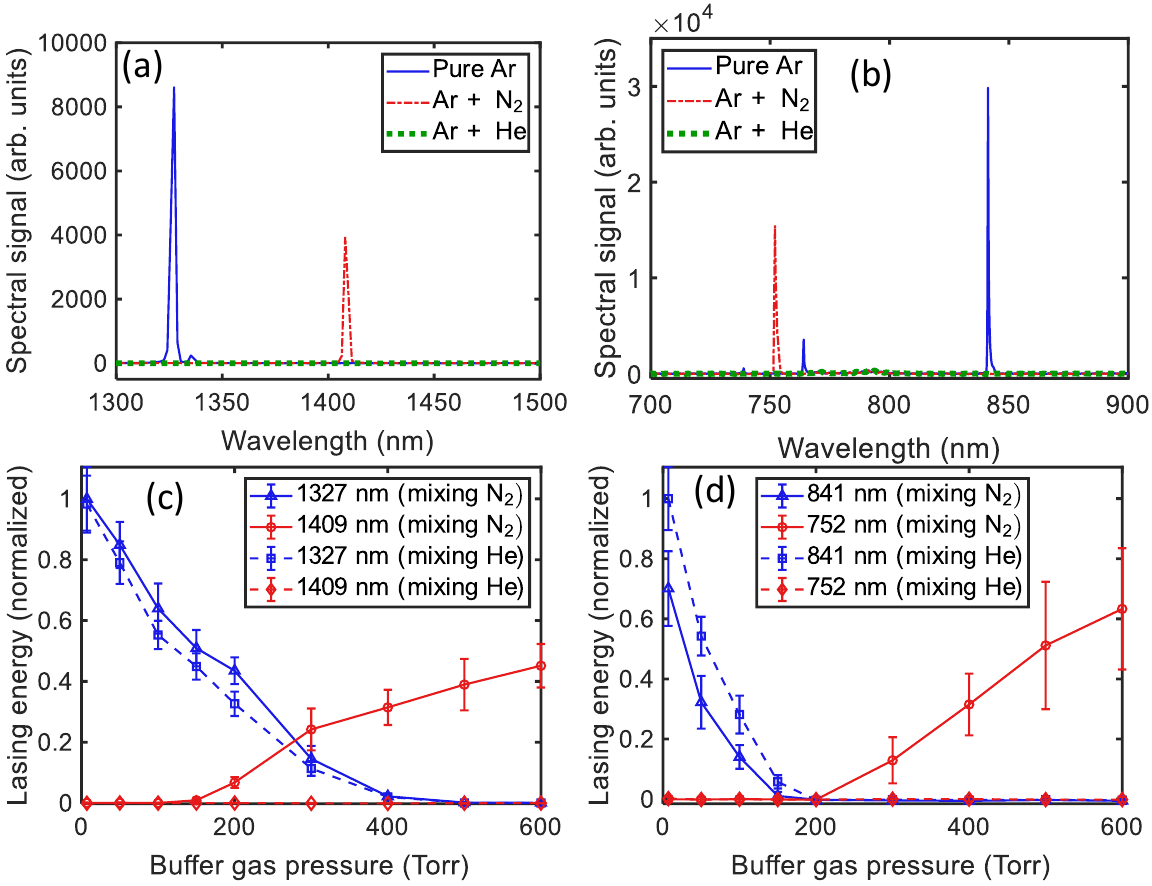}
\caption{\label{fig2} 
Evolution of forward cascade lasing spectra when mixing two different buffer gases (N$_2$ or He) into 7 Torr Ar gas. (a,b) 1st and 2nd lasing spectra in conditions of 7 Torr of pure Ar, 7 Torr of Ar mixed with 600 Torr of N$_2$, and 7 Torr of Ar mixed with 600 Torr of He. (c,d) Evolution of 1st and 2nd lasing energy when mixing different ratio of N$_2$ and He into 7 Torr Ar gas. 
}
\end{figure}

To study the origin of this switching, we added N$_2$ gas at various pressures to 7 Torr Ar in order to rule out the effect of other gases in air. The 1st and 2nd lasing spectra in conditions of 7 Torr of pure Ar, 7 Torr of Ar mixed with 600 Torr of N$_2$, and 7 Torr of Ar mixed with 600 Torr of helium (He) are shown in Fig. \ref{fig2}(a) and (b), respectively. Quantitatively (see solid blue and red lines in Fig. \ref{fig2}(c,d)), as increasing amount of N$_2$ is added to 7 Torr Ar gas, the 1st lasing wavelength switches from 1327.3 nm to 1409.4 nm and the 2nd lasing wavelength switches from 840.8 nm to 751.5 nm, similar to the case of atmospheric air. The detailed spectral data are shown in Supplementary Fig. 3. When He is added instead of N$_2$, the 1327.3 nm and 840.8 nm cascade is suppressed but the 1409.4 nm and the 751.5 nm signals do not appear, see dashed blue and red lines in Fig. \ref{fig2}(c) and (d). Thus, the 1409.4 nm and the 751.5 nm signals arise because of a new pathway enabled by N$_2$, while the 1327.3 nm and 840.8 nm signals are suppressed primarily by collisional relaxation with increasing density of He or N$_2$. As shown in Fig. \ref{fig1}(b), the dipole allowed transition between the ground and 3d[3/2]$_1$ states of Ar is resonant with the dipole transition in N$_2$ between its ground and electronically and vibrationally excited b$^{\prime 1}\Sigma_u^+$ (v=15) states, with the detuning below 2.6 meV. Coherence excited by three-photon absorption in N$_2$ leads to coherent macroscopic emission at the frequency resonant with the excitation of 3d[3/2]$_1$ states of Ar, exciting it from the ground state and triggering the 1409.4 nm and 751.5 nm cascade.

To check this hypothesis, we mixed Ar with varying concentrations of O$_2$, which is also one of the major constituents of air but lacks the required resonance for three-photon absorption of the 261 nm pump light. The observed spectral behavior is similar to that in the Ar-He mixture. This further supports our hypothesis of resonant energy transfer between N$_2$ and Ar as being responsible for the 1409.4 nm and 751.5 nm cascade lasing in atmospheric air. 

We quantified the three-photon absorption cross sections in N$_2$ and Ar by measuring absorption as a function of pump energy (see Supplementary Note for details). Our measured value of three-photon (nonlinear) absorption cross section in N$_2$ of $\sigma$(3) = $2.4\times 10^{-84}$ cm$^6$s$^2$ is 12 times smaller than the three-photon cross section of $\sigma$(3) = $3.0\times 10^{-83}$ cm$^6$s$^2$ in Ar. However, there is 78 times more N$_2$ than Ar in air, making three-photon excitation of N$_2$ dominant. 

\begin{figure}[tp]
\includegraphics[width=0.5\textwidth]{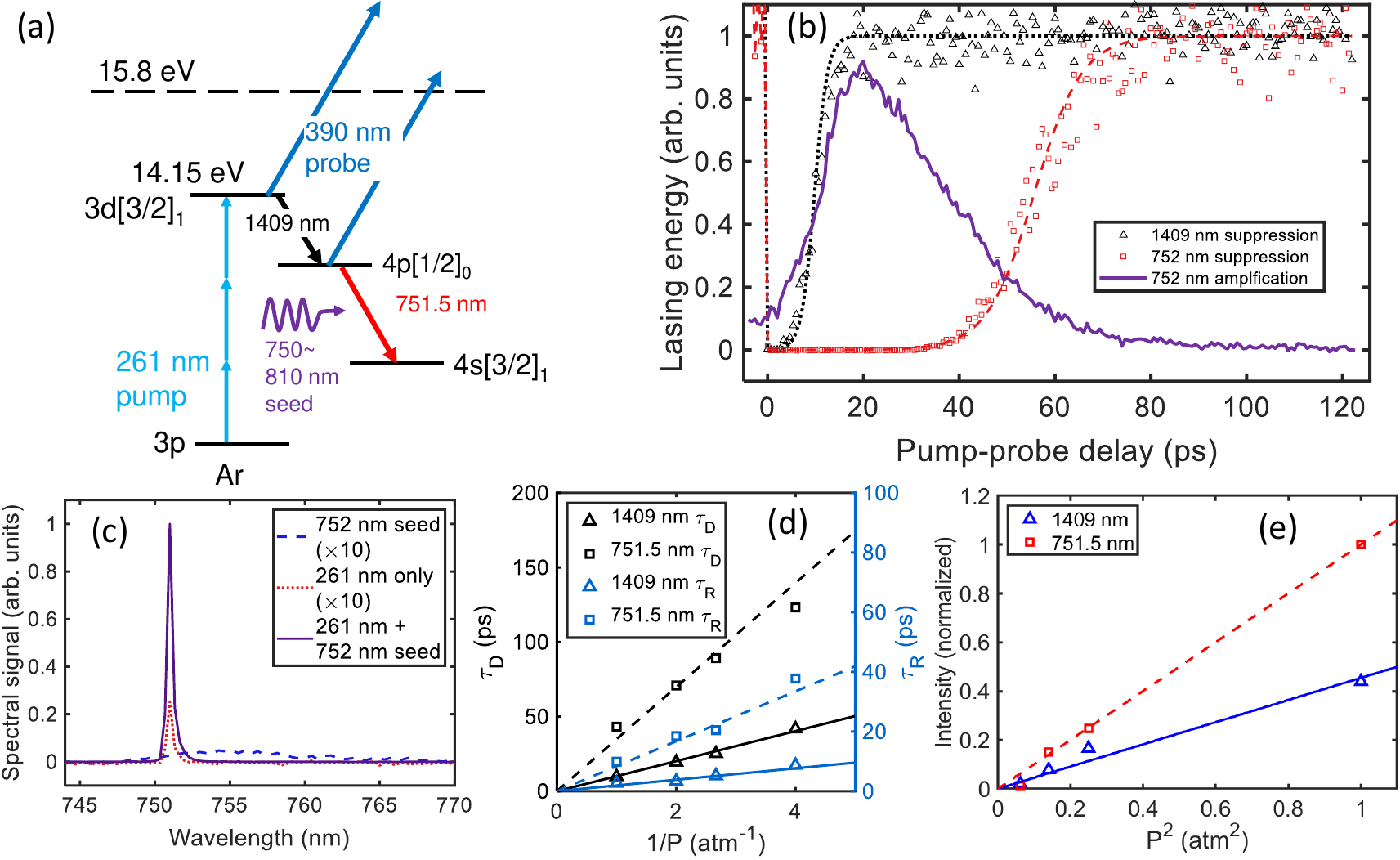}
\caption{\label{fig3} 
Pump-probe measurement results in air. (a) Diagram of relevant transitions of pump-probe experiment. (b) The suppression of 1st lasing (1409 nm, black) and 2nd lasing (752 nm, red), and the amplification of the seed pulse (purple) on 2nd lasing wavelength (752 nm) v.s. pump-probe delay in 1 atm air. The dotted black and dashed red curves are the best fit to a shifted tanh function. (c) The spectra of 2nd lasing with 752 nm seed only, 261 nm pump only, and both pump and seed (261 nm + 752 nm) at a delay time of 20 ps in atmospheric air. (d) Delay times and pulse durations of 1st and 2nd lasing signals as a function of the inverse of pressure. (e) Peak intensities of 1st and 2nd lasing signals compared to the square of pressure. Solid and dashed lines in (d) and (e) show a linear fit to measured data.
}
\end{figure}

We used the pump-probe approach to reveal the temporal dynamics of the forward two air lasing cascades, as shown in Fig. \ref{fig3}(a). Once the 261 nm pump pulse (light blue arrows in Fig. \ref{fig3}(a), energy 20 $\mu$J, peak intensity $4.0\times 10^{13}$ W/cm$^2$) excites the Ar atoms to the 3d[3/2]$_1$ manifold, the $\sim$390 nm probe pulse (blue arrows in Fig. \ref{fig3}(a), energy 60 $\mu$J, peak intensity $5.3\times 10^{13}$ W/cm$^2$) can one-photon ionize Ar atoms in both 3d and 4p states, leading to a reduction of gain for both lasing transitions. Figure \ref{fig3}(b) shows the 1st lasing (black triangle mark) and 2nd lasing (red square mark) signals as a function of pump-probe delay for a 390 nm probe in 1 atm air. When the probe pulse arrives before the pump, the 1409 nm lasing signal is not suppressed because the probe does not affect the lasing process. When the probe arrives just after the pump, both the 1st and 2nd lasing signals are strongly suppressed due to depletion of the 3d and 4p state via single-photon ionization by the probe. When the probe pulse is delayed about 20 ps later than the pump, the 1st lasing signal recovers to its original level, which means that the 1st lasing signal is emitted within $\sim$20 ps after the pump. Similarly, the 2nd lasing signal is emitted within $\sim$80 ps after the pump. The delay-dependent signals in Fig. \ref{fig3}(b) after the maximal suppression delay are proportional to the time integral of the energy emitted in the 1st and 2nd lasing steps until the arrival of the probe. Therefore, we can retreive the pulse profile of the 1st and 2nd lasing and accordingly their delay time $\tau_D$ and pulse durations $\tau_R$ by taking the derivatives of the recovery time of the curves. The dotted black and dashed red curves in Fig. \ref{fig3}(b) show the best fit of the measured data to a hyperbolic tangent function, whose derivative is a sech$^2$ pulse. The delay time ($\tau_D$) and pulse width ($\tau_R$) for both the 1st and 2nd lasing pulse at pressures of 1.0, 0.5, 0.375, and 0.25 atm are shown in Fig. \ref{fig3}(d). All of them are inversely proportional to the air pressure. From the measured energy, pulse duration, and spot size, we obtain the peak intensities of the SF as shown in Fig. \ref{fig3}(e), which are proportional to the square of air pressure. Pulse width and delay time being inversely proportioinal to density and lasing intensity being proportional to the square of the density are the two key signatures of SF\cite{maki_influence_1989,kitano_cascade_2023}. Therefore, we confirm that both the 1st and 2nd lasing in air arise from SF. Since the 2nd lasing is delayed respective to the 1st lasing, they are cascade SF instead of yoked SF\cite{brownell_yoked_1995}.

Furthermore, by using a weak ($<0.1~\mu$J, peak intensity $<2.2\times 10^{10}$ W/cm$^2$) seed pulse centered at 780 nm with sufficient bandwidth covering the spectrum of the 2nd lasing, we can directly see the amplification of the 2nd lasing signal (purple curve in Fig. \ref{fig3}(b) and (c)). The seed is too weak to suppress the 1st lasing signal, but can amplify the 2nd lasing signal once the 4p[1/2]$_0$ level is populated. The rise and fall of this signal reflect the rise and fall of population inversion in the 2nd lasing step. Figure \ref{fig3}(c) shows the spectra of the 2nd lasing with 752 nm seed only, 261 nm pump only, and both pump and seed (261 nm + 752 nm) at a delay time of 20 ps in atmospheric air. At this delay time, the amplification factor reaches over 100, showing a high optical gain of the 2nd lasing. As expected, the spectrum of the amplified pulse is very narrow compared to that of the seed pulse.

From Fig. \ref{fig3}(b) and (d), we can see that the delay time of the 1409 nm 1st lasing pulse in 1 atm air is $\sim$10 ps. When similar measurements are performed under identical experimental conditions but in 7 Torr of pure Ar, the delay of the dominant 1327 nm 1st lasing pulse is $\sim$30 ps (see Supplementary Fig. 5). This means that the gain of the 1409 nm lasing in atmospheric air is much larger than that of the 1327 nm lasing in 7 Torr of pure Ar. Also, the three times longer build-up time makes the 1327 nm lasing pathway more susceptible to collisional relaxation processes when buffer gases are added\cite{maki_influence_1989}. The collision rate between excited Ar atoms and N$_2$ molecules can be estimated using the formula\cite{verdeyen_laser_1995} $\nu_{coll}=N\sigma[\frac{8kT}{\pi} (\frac{1}{M_{Ar}} + \frac{1}{M_{N_2}} )]^{1/2}$. The cross section was calculated as $\sigma=\pi(r_{Ar}+r_{N_2} )^2$, where $r_{Ar}\approx n^{*2} a_0$, with $n^*$ = 2.99 is the effective quantum number of the 3d$^\prime$ excited state and $a_0$ is the Bohr radius, and $r_{N_2}\approx$ 185 pm is the radius of the N$_2$ molecule\cite{kunze_molecular_2022}. This yields an estimated collision time (inverse of collision rate) of 32 ps in 7 Torr of Ar mixed with 600 Torr of N$_2$, and 25 ps in 7 Torr of Ar mixed with 600 Torr of He, both of which are roughly the same as the delay time of the 1327 nm lasing signal in 7 Torr of pure Ar ($\sim$30 ps). This is the reason why the 1327 nm lasing signal quenches in atmospheric air.

Dipole selection rules\cite{grynberg_three-photon_1979} allow three-photon transitions from states with total angular momentum J=0 (e.g. Ar ground state) to states with either J=1 (e.g. 3d[3/2]$_1$) or J=3 (e.g. 3d$^\prime$[5/2]$_3$). However, transitions with $\Delta$J = 1 also have an allowed single-photon excitation pathway. In pure Ar gas at pressures above 1 Torr, excitation to the 3d[3/2]$_1$ state is suppressed by interference between three-photon absorption of pump light and single-photon absorption of third harmonic light\cite{jackson_interference_1982, payne_effects_1980}. This prevents 1409 nm emission and leaves 1327 nm as the most prominent signal, because the third harmonic single-photon absorption pathway to the 3d$^\prime$[5/2]$_3$ level is forbidden ($\Delta$J = 3) by dipole selection rules. In air, there is a large concentration of N$_2$, which has a dipole transition from the electronically and vibrationally excited b$^{\prime 1}\Sigma_u^+$ (v=15) state to ground with an energy defect of $<$2.6 meV compared to the energy of the dipole transition from 3d[3/2]$_1$ to ground in Ar. At frequencies within the linewidth of the single-photon Ar transition, the refractive index is usually dominated by the contribution from the Ar resonance. In air however, the large concentration of N$_2$ significantly modifies the refractive index at these frequencies, creating a phase matching condition which enhances third-harmonic generation (THG). The enhanced THG disrupts the balance between the single-photon and three-photon excitation pathways, enabling excitation to the 3d[3/2]$_1$ state. This larger population of the 3d[3/2]$_1$ state allows lasing on the 1409 nm transition to build up more quickly than on the 1327 nm transition and persist even in the presence of collisions in atmospheric air.

\begin{figure}[tp]
\includegraphics[width=0.4\textwidth]{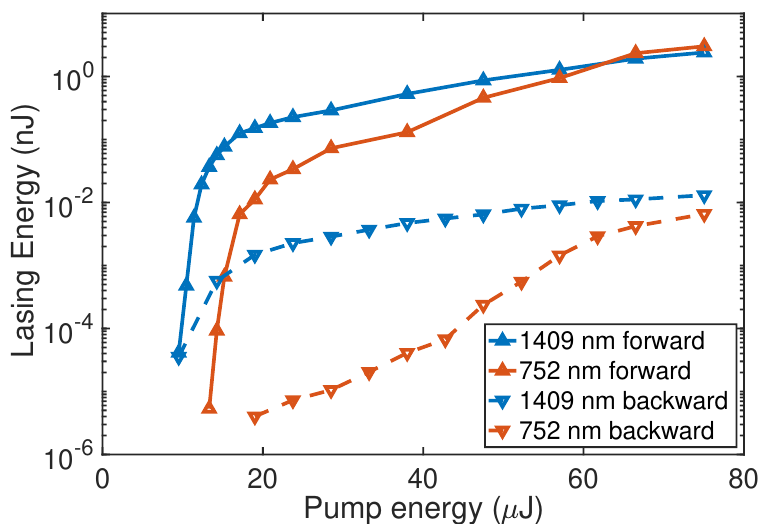}
\caption{\label{fig4} 
Comparison of forward and backward cascade superfluorescent lasing energy v.s. pump energy in atmospheric air.
}
\end{figure}

In atmospheric remote sensing applications, it is the backward lasing that enables single-ended standoff diagnostics. Here, we compare the cascade superfluorescent lasing energy in forward and backward directions. The blue and red solid (dashed) line in Fig. \ref{fig4} shows the 1409 nm 1st lasing energy and the 752 nm 2nd lasing energy versus pump energy in the forward (backward) direction, respectively. The error bar is smaller than the marker in the curves. For forward lasing, when the pump energy is lower than 18 $\mu$J, both the 1st and 2nd lasing show exponential growth with the increase of pump energy, indicating ASE at low pump intensities. When the pump energy exceeds 18 $\mu$J, both the 1st and 2nd lasing energy show a roughly quadratic relation with the pump energy and as seen and inferred from Fig. \ref{fig3}(d) and (e), the emission transits from ASE to SF. Ideally, for SF emission, the lasing energy should grow as the third power of the pump energy in absence of losses. Additional experimentation is needed to understand the discrepancy. At highest pump energy, the forward (backward) 1st and 2nd lasing efficiency respective to the pump energy is 3.2$\times 10^{-5}$ and 4.0$\times 10^{-5}$ (1.7$\times 10^{-7}$ and 0.9$\times 10^{-7}$), respectively. The backward lasing energy is about two orders of magnitude lower than the forward lasing energy due to the nature of travelling wave excitation in the pumping process, given that the effective gain lifetime ($\sim$10 ps) is much shorter than the propagation time over a $\sim$1-cm-long excitation volume. Previous studies\cite{dogariu_high-gain_2011,laurain_low-threshold_2014, dogariu_three-photon_2016} also showed that backward lasing signal can be even stronger than the forward lasing signal by using longer (nanosecond) pump pulses, which is also confirmed by our simulation results by numerically solving the Maxwell-Bloch equations. In this way, the pump pulse duration is much longer than the propagation time of the excitation volume so that travelling wave excitation effect can be ignored.

We have analyzed potential issue of attenuation of the 261 nm radiation propagating through atmosphere for remote standoff detection at long distances. Our calculations, using normal atmosphere and HITRAN data\cite{gordon_hitran2020_2022} show that attenuation of 261 nm radiation, including losses and scattering, is approximately 25\% for 1 km propagation, i.e., $\sim$75\% of the 261 nm pulse energy at low intensity will be delivered at a distance of 1 km. Incidentally, majority of the attenuation arises from Raleigh scattering.

In conclusion, we have demonstrated bidirectional two-color cascade superfluorescent lasing in Ar in atmospheric air pumped by femtosecond UV pulses at 261 nm and uncovered photon-mediated resonant energy transfer from N$_2$ to Ar. Time-resolved measurements confirmed that the cascaded lasing arises from SF. 

\begin{acknowledgments}
We thank Sergei Tochitsky for useful discussions regarding this work. This work was supported by DOE Award No. DE-SC0010064, NSF grant No. 2003354, and the Office of Naval Research (ONR) Multidisciplinary University Research Initiative (MURI) No. 14-19-1-2517. The computation is completed in the HPC Platform of Huazhong University of Science and Technology.
\end{acknowledgments}


%\bibliography{refs}

\begin{thebibliography}{41}%
\makeatletter
\providecommand \@ifxundefined [1]{%
 \@ifx{#1\undefined}
}%
\providecommand \@ifnum [1]{%
 \ifnum #1\expandafter \@firstoftwo
 \else \expandafter \@secondoftwo
 \fi
}%
\providecommand \@ifx [1]{%
 \ifx #1\expandafter \@firstoftwo
 \else \expandafter \@secondoftwo
 \fi
}%
\providecommand \natexlab [1]{#1}%
\providecommand \enquote  [1]{``#1''}%
\providecommand \bibnamefont  [1]{#1}%
\providecommand \bibfnamefont [1]{#1}%
\providecommand \citenamefont [1]{#1}%
\providecommand \href@noop [0]{\@secondoftwo}%
\providecommand \href [0]{\begingroup \@sanitize@url \@href}%
\providecommand \@href[1]{\@@startlink{#1}\@@href}%
\providecommand \@@href[1]{\endgroup#1\@@endlink}%
\providecommand \@sanitize@url [0]{\catcode `\\12\catcode `\$12\catcode
  `\&12\catcode `\#12\catcode `\^12\catcode `\_12\catcode `\%12\relax}%
\providecommand \@@startlink[1]{}%
\providecommand \@@endlink[0]{}%
\providecommand \url  [0]{\begingroup\@sanitize@url \@url }%
\providecommand \@url [1]{\endgroup\@href {#1}{\urlprefix }}%
\providecommand \urlprefix  [0]{URL }%
\providecommand \Eprint [0]{\href }%
\providecommand \doibase [0]{https://doi.org/}%
\providecommand \selectlanguage [0]{\@gobble}%
\providecommand \bibinfo  [0]{\@secondoftwo}%
\providecommand \bibfield  [0]{\@secondoftwo}%
\providecommand \translation [1]{[#1]}%
\providecommand \BibitemOpen [0]{}%
\providecommand \bibitemStop [0]{}%
\providecommand \bibitemNoStop [0]{.\EOS\space}%
\providecommand \EOS [0]{\spacefactor3000\relax}%
\providecommand \BibitemShut  [1]{\csname bibitem#1\endcsname}%
\let\auto@bib@innerbib\@empty
%</preamble>
\bibitem [{\citenamefont {Laurain}\ \emph {et~al.}(2014)\citenamefont
  {Laurain}, \citenamefont {Scheller},\ and\ \citenamefont
  {Polynkin}}]{laurain_low-threshold_2014}%
  \BibitemOpen
  \bibfield  {author} {\bibinfo {author} {\bibfnamefont {A.}~\bibnamefont
  {Laurain}}, \bibinfo {author} {\bibfnamefont {M.}~\bibnamefont {Scheller}},\
  and\ \bibinfo {author} {\bibfnamefont {P.}~\bibnamefont {Polynkin}},\
  }\bibfield  {title} {\bibinfo {title} {Low-{Threshold} {Bidirectional} {Air}
  {Lasing}},\ }\href {https://doi.org/10.1103/PhysRevLett.113.253901}
  {\bibfield  {journal} {\bibinfo  {journal} {Phys. Rev. Lett.}\ }\textbf
  {\bibinfo {volume} {113}},\ \bibinfo {pages} {253901} (\bibinfo {year}
  {2014})}\BibitemShut {NoStop}%
\bibitem [{\citenamefont {Dogariu}\ \emph {et~al.}(2011)\citenamefont
  {Dogariu}, \citenamefont {Michael}, \citenamefont {Scully},\ and\
  \citenamefont {Miles}}]{dogariu_high-gain_2011}%
  \BibitemOpen
  \bibfield  {author} {\bibinfo {author} {\bibfnamefont {A.}~\bibnamefont
  {Dogariu}}, \bibinfo {author} {\bibfnamefont {J.~B.}\ \bibnamefont
  {Michael}}, \bibinfo {author} {\bibfnamefont {M.~O.}\ \bibnamefont
  {Scully}},\ and\ \bibinfo {author} {\bibfnamefont {R.~B.}\ \bibnamefont
  {Miles}},\ }\bibfield  {title} {\bibinfo {title} {High-{Gain} {Backward}
  {Lasing} in {Air}},\ }\href {https://doi.org/10.1126/science.1199492}
  {\bibfield  {journal} {\bibinfo  {journal} {Science}\ }\textbf {\bibinfo
  {volume} {331}},\ \bibinfo {pages} {442} (\bibinfo {year}
  {2011})}\BibitemShut {NoStop}%
\bibitem [{\citenamefont {Hemmer}\ \emph {et~al.}(2011)\citenamefont {Hemmer},
  \citenamefont {Miles}, \citenamefont {Polynkin}, \citenamefont {Siebert},
  \citenamefont {Sokolov}, \citenamefont {Sprangle},\ and\ \citenamefont
  {Scully}}]{hemmer_standoff_2011}%
  \BibitemOpen
  \bibfield  {author} {\bibinfo {author} {\bibfnamefont {P.~R.}\ \bibnamefont
  {Hemmer}}, \bibinfo {author} {\bibfnamefont {R.~B.}\ \bibnamefont {Miles}},
  \bibinfo {author} {\bibfnamefont {P.}~\bibnamefont {Polynkin}}, \bibinfo
  {author} {\bibfnamefont {T.}~\bibnamefont {Siebert}}, \bibinfo {author}
  {\bibfnamefont {A.~V.}\ \bibnamefont {Sokolov}}, \bibinfo {author}
  {\bibfnamefont {P.}~\bibnamefont {Sprangle}},\ and\ \bibinfo {author}
  {\bibfnamefont {M.~O.}\ \bibnamefont {Scully}},\ }\bibfield  {title}
  {\bibinfo {title} {Standoff spectroscopy via remote generation of a
  backward-propagating laser beam},\ }\href
  {https://doi.org/10.1073/pnas.1014401107} {\bibfield  {journal} {\bibinfo
  {journal} {Proc. Natl. Acad. Sci. U.S.A.}\ }\textbf {\bibinfo {volume}
  {108}},\ \bibinfo {pages} {3130} (\bibinfo {year} {2011})}\BibitemShut
  {NoStop}%
\bibitem [{\citenamefont {Malevich}\ \emph {et~al.}(2015)\citenamefont
  {Malevich}, \citenamefont {Maurer}, \citenamefont {Kartashov}, \citenamefont
  {Ališauskas}, \citenamefont {Lanin}, \citenamefont {Zheltikov},
  \citenamefont {Marangoni}, \citenamefont {Cerullo}, \citenamefont
  {Baltuška},\ and\ \citenamefont {Pugžlys}}]{malevich_stimulated_2015}%
  \BibitemOpen
  \bibfield  {author} {\bibinfo {author} {\bibfnamefont {P.~N.}\ \bibnamefont
  {Malevich}}, \bibinfo {author} {\bibfnamefont {R.}~\bibnamefont {Maurer}},
  \bibinfo {author} {\bibfnamefont {D.}~\bibnamefont {Kartashov}}, \bibinfo
  {author} {\bibfnamefont {S.}~\bibnamefont {Ališauskas}}, \bibinfo {author}
  {\bibfnamefont {A.~A.}\ \bibnamefont {Lanin}}, \bibinfo {author}
  {\bibfnamefont {A.~M.}\ \bibnamefont {Zheltikov}}, \bibinfo {author}
  {\bibfnamefont {M.}~\bibnamefont {Marangoni}}, \bibinfo {author}
  {\bibfnamefont {G.}~\bibnamefont {Cerullo}}, \bibinfo {author} {\bibfnamefont
  {A.}~\bibnamefont {Baltuška}},\ and\ \bibinfo {author} {\bibfnamefont
  {A.}~\bibnamefont {Pugžlys}},\ }\bibfield  {title} {\bibinfo {title}
  {Stimulated {Raman} gas sensing by backward {UV} lasing from a femtosecond
  filament},\ }\href {https://doi.org/10.1364/OL.40.002469} {\bibfield
  {journal} {\bibinfo  {journal} {Opt. Lett.}\ }\textbf {\bibinfo {volume}
  {40}},\ \bibinfo {pages} {2469} (\bibinfo {year} {2015})}\BibitemShut
  {NoStop}%
\bibitem [{\citenamefont {Zhang}\ \emph {et~al.}(2022)\citenamefont {Zhang},
  \citenamefont {Zhang}, \citenamefont {Xu}, \citenamefont {Xie}, \citenamefont
  {Fu}, \citenamefont {Lu}, \citenamefont {Zhang}, \citenamefont {Yu},
  \citenamefont {Yao}, \citenamefont {Cheng},\ and\ \citenamefont
  {Xu}}]{zhang_high-sensitivity_2022}%
  \BibitemOpen
  \bibfield  {author} {\bibinfo {author} {\bibfnamefont {Z.}~\bibnamefont
  {Zhang}}, \bibinfo {author} {\bibfnamefont {F.}~\bibnamefont {Zhang}},
  \bibinfo {author} {\bibfnamefont {B.}~\bibnamefont {Xu}}, \bibinfo {author}
  {\bibfnamefont {H.}~\bibnamefont {Xie}}, \bibinfo {author} {\bibfnamefont
  {B.}~\bibnamefont {Fu}}, \bibinfo {author} {\bibfnamefont {X.}~\bibnamefont
  {Lu}}, \bibinfo {author} {\bibfnamefont {N.}~\bibnamefont {Zhang}}, \bibinfo
  {author} {\bibfnamefont {S.}~\bibnamefont {Yu}}, \bibinfo {author}
  {\bibfnamefont {J.}~\bibnamefont {Yao}}, \bibinfo {author} {\bibfnamefont
  {Y.}~\bibnamefont {Cheng}},\ and\ \bibinfo {author} {\bibfnamefont
  {Z.}~\bibnamefont {Xu}},\ }\bibfield  {title} {\bibinfo {title}
  {High-{Sensitivity} {Gas} {Detection} with {Air}-{Lasing}-{Assisted}
  {Coherent} {Raman} {Spectroscopy}},\ }\href
  {https://doi.org/10.34133/2022/9761458} {\bibfield  {journal} {\bibinfo
  {journal} {Ultrafast Sci}\ }\textbf {\bibinfo {volume} {2022}},\ \bibinfo
  {pages} {9761458} (\bibinfo {year} {2022})}\BibitemShut {NoStop}%
\bibitem [{\citenamefont {Fu}\ \emph {et~al.}(2022)\citenamefont {Fu},
  \citenamefont {Cao}, \citenamefont {Yamanouchi},\ and\ \citenamefont
  {Xu}}]{fu_air-laser-based_2022}%
  \BibitemOpen
  \bibfield  {author} {\bibinfo {author} {\bibfnamefont {Y.}~\bibnamefont
  {Fu}}, \bibinfo {author} {\bibfnamefont {J.}~\bibnamefont {Cao}}, \bibinfo
  {author} {\bibfnamefont {K.}~\bibnamefont {Yamanouchi}},\ and\ \bibinfo
  {author} {\bibfnamefont {H.}~\bibnamefont {Xu}},\ }\bibfield  {title}
  {\bibinfo {title} {Air-{Laser}-{Based} {Standoff} {Coherent} {Raman}
  {Spectrometer}},\ }\href {https://doi.org/10.34133/2022/9867028} {\bibfield
  {journal} {\bibinfo  {journal} {Ultrafast Sci}\ }\textbf {\bibinfo {volume}
  {2022}},\ \bibinfo {pages} {9867028} (\bibinfo {year} {2022})}\BibitemShut
  {NoStop}%
\bibitem [{\citenamefont {Zhang}\ \emph {et~al.}(2023)\citenamefont {Zhang},
  \citenamefont {Xie}, \citenamefont {Zhang}, \citenamefont {Lu}, \citenamefont
  {Chen}, \citenamefont {Wu}, \citenamefont {Cheng},\ and\ \citenamefont
  {Yao}}]{zhang_electronicresonanceenhanced_2023}%
  \BibitemOpen
  \bibfield  {author} {\bibinfo {author} {\bibfnamefont {N.}~\bibnamefont
  {Zhang}}, \bibinfo {author} {\bibfnamefont {H.}~\bibnamefont {Xie}}, \bibinfo
  {author} {\bibfnamefont {H.}~\bibnamefont {Zhang}}, \bibinfo {author}
  {\bibfnamefont {X.}~\bibnamefont {Lu}}, \bibinfo {author} {\bibfnamefont
  {Y.}~\bibnamefont {Chen}}, \bibinfo {author} {\bibfnamefont {Y.}~\bibnamefont
  {Wu}}, \bibinfo {author} {\bibfnamefont {Y.}~\bibnamefont {Cheng}},\ and\
  \bibinfo {author} {\bibfnamefont {J.}~\bibnamefont {Yao}},\ }\bibfield
  {title} {\bibinfo {title} {Electronic‐{Resonance}‐{Enhanced} {Coherent}
  {Raman} {Spectroscopy} with a {Single} {Femtosecond} {Laser} {Beam}},\ }\href
  {https://doi.org/10.1002/lpor.202300020} {\bibfield  {journal} {\bibinfo
  {journal} {Laser \& Photonics Reviews}\ ,\ \bibinfo {pages} {2300020}}
  (\bibinfo {year} {2023})}\BibitemShut {NoStop}%
\bibitem [{\citenamefont {Yao}\ \emph {et~al.}(2011)\citenamefont {Yao},
  \citenamefont {Zeng}, \citenamefont {Xu}, \citenamefont {Li}, \citenamefont
  {Chu}, \citenamefont {Ni}, \citenamefont {Zhang}, \citenamefont {Chin},
  \citenamefont {Cheng},\ and\ \citenamefont {Xu}}]{yao_high-brightness_2011}%
  \BibitemOpen
  \bibfield  {author} {\bibinfo {author} {\bibfnamefont {J.}~\bibnamefont
  {Yao}}, \bibinfo {author} {\bibfnamefont {B.}~\bibnamefont {Zeng}}, \bibinfo
  {author} {\bibfnamefont {H.}~\bibnamefont {Xu}}, \bibinfo {author}
  {\bibfnamefont {G.}~\bibnamefont {Li}}, \bibinfo {author} {\bibfnamefont
  {W.}~\bibnamefont {Chu}}, \bibinfo {author} {\bibfnamefont {J.}~\bibnamefont
  {Ni}}, \bibinfo {author} {\bibfnamefont {H.}~\bibnamefont {Zhang}}, \bibinfo
  {author} {\bibfnamefont {S.~L.}\ \bibnamefont {Chin}}, \bibinfo {author}
  {\bibfnamefont {Y.}~\bibnamefont {Cheng}},\ and\ \bibinfo {author}
  {\bibfnamefont {Z.}~\bibnamefont {Xu}},\ }\bibfield  {title} {\bibinfo
  {title} {High-brightness switchable multiwavelength remote laser in air},\
  }\href {https://doi.org/10.1103/PhysRevA.84.051802} {\bibfield  {journal}
  {\bibinfo  {journal} {Phys. Rev. A}\ }\textbf {\bibinfo {volume} {84}},\
  \bibinfo {pages} {051802(R)} (\bibinfo {year} {2011})}\BibitemShut {NoStop}%
\bibitem [{\citenamefont {Zhang}\ \emph {et~al.}(2013)\citenamefont {Zhang},
  \citenamefont {Jing}, \citenamefont {Yao}, \citenamefont {Li}, \citenamefont
  {Zeng}, \citenamefont {Chu}, \citenamefont {Ni}, \citenamefont {Xie},
  \citenamefont {Xu}, \citenamefont {Chin}, \citenamefont {Yamanouchi},
  \citenamefont {Cheng},\ and\ \citenamefont {Xu}}]{zhang_rotational_2013}%
  \BibitemOpen
  \bibfield  {author} {\bibinfo {author} {\bibfnamefont {H.}~\bibnamefont
  {Zhang}}, \bibinfo {author} {\bibfnamefont {C.}~\bibnamefont {Jing}},
  \bibinfo {author} {\bibfnamefont {J.}~\bibnamefont {Yao}}, \bibinfo {author}
  {\bibfnamefont {G.}~\bibnamefont {Li}}, \bibinfo {author} {\bibfnamefont
  {B.}~\bibnamefont {Zeng}}, \bibinfo {author} {\bibfnamefont {W.}~\bibnamefont
  {Chu}}, \bibinfo {author} {\bibfnamefont {J.}~\bibnamefont {Ni}}, \bibinfo
  {author} {\bibfnamefont {H.}~\bibnamefont {Xie}}, \bibinfo {author}
  {\bibfnamefont {H.}~\bibnamefont {Xu}}, \bibinfo {author} {\bibfnamefont
  {S.~L.}\ \bibnamefont {Chin}}, \bibinfo {author} {\bibfnamefont
  {K.}~\bibnamefont {Yamanouchi}}, \bibinfo {author} {\bibfnamefont
  {Y.}~\bibnamefont {Cheng}},\ and\ \bibinfo {author} {\bibfnamefont
  {Z.}~\bibnamefont {Xu}},\ }\bibfield  {title} {\bibinfo {title} {Rotational
  {Coherence} {Encoded} in an “{Air}-{Laser}” {Spectrum} of {Nitrogen}
  {Molecular} {Ions} in an {Intense} {Laser} {Field}},\ }\href
  {https://doi.org/10.1103/PhysRevX.3.041009} {\bibfield  {journal} {\bibinfo
  {journal} {Phys. Rev. X}\ }\textbf {\bibinfo {volume} {3}},\ \bibinfo {pages}
  {041009} (\bibinfo {year} {2013})}\BibitemShut {NoStop}%
\bibitem [{\citenamefont {Liu}\ \emph {et~al.}(2015)\citenamefont {Liu},
  \citenamefont {Ding}, \citenamefont {Lambert}, \citenamefont {Houard},
  \citenamefont {Tikhonchuk},\ and\ \citenamefont
  {Mysyrowicz}}]{liu_recollision-induced_2015}%
  \BibitemOpen
  \bibfield  {author} {\bibinfo {author} {\bibfnamefont {Y.}~\bibnamefont
  {Liu}}, \bibinfo {author} {\bibfnamefont {P.}~\bibnamefont {Ding}}, \bibinfo
  {author} {\bibfnamefont {G.}~\bibnamefont {Lambert}}, \bibinfo {author}
  {\bibfnamefont {A.}~\bibnamefont {Houard}}, \bibinfo {author} {\bibfnamefont
  {V.}~\bibnamefont {Tikhonchuk}},\ and\ \bibinfo {author} {\bibfnamefont
  {A.}~\bibnamefont {Mysyrowicz}},\ }\bibfield  {title} {\bibinfo {title}
  {Recollision-{Induced} {Superradiance} of {Ionized} {Nitrogen} {Molecules}},\
  }\href {https://doi.org/10.1103/PhysRevLett.115.133203} {\bibfield  {journal}
  {\bibinfo  {journal} {Phys. Rev. Lett.}\ }\textbf {\bibinfo {volume} {115}},\
  \bibinfo {pages} {133203} (\bibinfo {year} {2015})}\BibitemShut {NoStop}%
\bibitem [{\citenamefont {Xu}\ \emph {et~al.}(2015)\citenamefont {Xu},
  \citenamefont {Lötstedt}, \citenamefont {Iwasaki},\ and\ \citenamefont
  {Yamanouchi}}]{xu_sub-10-fs_2015}%
  \BibitemOpen
  \bibfield  {author} {\bibinfo {author} {\bibfnamefont {H.}~\bibnamefont
  {Xu}}, \bibinfo {author} {\bibfnamefont {E.}~\bibnamefont {Lötstedt}},
  \bibinfo {author} {\bibfnamefont {A.}~\bibnamefont {Iwasaki}},\ and\ \bibinfo
  {author} {\bibfnamefont {K.}~\bibnamefont {Yamanouchi}},\ }\bibfield  {title}
  {\bibinfo {title} {Sub-10-fs population inversion in {N}$_2^+$ in air lasing
  through multiple state coupling},\ }\href
  {https://doi.org/10.1038/ncomms9347} {\bibfield  {journal} {\bibinfo
  {journal} {Nat Commun}\ }\textbf {\bibinfo {volume} {6}},\ \bibinfo {pages}
  {8347} (\bibinfo {year} {2015})}\BibitemShut {NoStop}%
\bibitem [{\citenamefont {Dogariu}\ and\ \citenamefont
  {Miles}(2016)}]{dogariu_three-photon_2016}%
  \BibitemOpen
  \bibfield  {author} {\bibinfo {author} {\bibfnamefont {A.}~\bibnamefont
  {Dogariu}}\ and\ \bibinfo {author} {\bibfnamefont {R.~B.}\ \bibnamefont
  {Miles}},\ }\bibfield  {title} {\bibinfo {title} {Three-photon femtosecond
  pumped backwards lasing in argon},\ }\href
  {https://doi.org/10.1364/OE.24.00A544} {\bibfield  {journal} {\bibinfo
  {journal} {Opt. Express}\ }\textbf {\bibinfo {volume} {24}},\ \bibinfo
  {pages} {A544} (\bibinfo {year} {2016})}\BibitemShut {NoStop}%
\bibitem [{\citenamefont {Yao}\ \emph {et~al.}(2016)\citenamefont {Yao},
  \citenamefont {Jiang}, \citenamefont {Chu}, \citenamefont {Zeng},
  \citenamefont {Wu}, \citenamefont {Lu}, \citenamefont {Li}, \citenamefont
  {Xie}, \citenamefont {Li}, \citenamefont {Yu}, \citenamefont {Wang},
  \citenamefont {Jiang}, \citenamefont {Gong},\ and\ \citenamefont
  {Cheng}}]{yao_population_2016}%
  \BibitemOpen
  \bibfield  {author} {\bibinfo {author} {\bibfnamefont {J.}~\bibnamefont
  {Yao}}, \bibinfo {author} {\bibfnamefont {S.}~\bibnamefont {Jiang}}, \bibinfo
  {author} {\bibfnamefont {W.}~\bibnamefont {Chu}}, \bibinfo {author}
  {\bibfnamefont {B.}~\bibnamefont {Zeng}}, \bibinfo {author} {\bibfnamefont
  {C.}~\bibnamefont {Wu}}, \bibinfo {author} {\bibfnamefont {R.}~\bibnamefont
  {Lu}}, \bibinfo {author} {\bibfnamefont {Z.}~\bibnamefont {Li}}, \bibinfo
  {author} {\bibfnamefont {H.}~\bibnamefont {Xie}}, \bibinfo {author}
  {\bibfnamefont {G.}~\bibnamefont {Li}}, \bibinfo {author} {\bibfnamefont
  {C.}~\bibnamefont {Yu}}, \bibinfo {author} {\bibfnamefont {Z.}~\bibnamefont
  {Wang}}, \bibinfo {author} {\bibfnamefont {H.}~\bibnamefont {Jiang}},
  \bibinfo {author} {\bibfnamefont {Q.}~\bibnamefont {Gong}},\ and\ \bibinfo
  {author} {\bibfnamefont {Y.}~\bibnamefont {Cheng}},\ }\bibfield  {title}
  {\bibinfo {title} {Population {Redistribution} {Among} {Multiple}
  {Electronic} {States} of {Molecular} {Nitrogen} {Ions} in {Strong} {Laser}
  {Fields}},\ }\href {https://doi.org/10.1103/PhysRevLett.116.143007}
  {\bibfield  {journal} {\bibinfo  {journal} {Phys. Rev. Lett.}\ }\textbf
  {\bibinfo {volume} {116}},\ \bibinfo {pages} {143007} (\bibinfo {year}
  {2016})}\BibitemShut {NoStop}%
\bibitem [{\citenamefont {Polynkin}\ and\ \citenamefont
  {Cheng}(2018)}]{polynkin_air_2018}%
  \BibitemOpen
  \bibinfo {editor} {\bibfnamefont {P.}~\bibnamefont {Polynkin}}\ and\ \bibinfo
  {editor} {\bibfnamefont {Y.}~\bibnamefont {Cheng}},\ eds.,\ \href
  {https://doi.org/10.1007/978-3-319-65220-7} {\emph {\bibinfo {title} {Air
  {Lasing}}}},\ \bibinfo {series} {Springer {Series} in {Optical} {Sciences}},
  Vol.\ \bibinfo {volume} {208}\ (\bibinfo  {publisher} {Springer International
  Publishing},\ \bibinfo {address} {Cham},\ \bibinfo {year} {2018})\BibitemShut
  {NoStop}%
\bibitem [{\citenamefont {Yuan}\ \emph {et~al.}(2019)\citenamefont {Yuan},
  \citenamefont {Liu}, \citenamefont {Yao},\ and\ \citenamefont
  {Cheng}}]{yuan_recent_2019}%
  \BibitemOpen
  \bibfield  {author} {\bibinfo {author} {\bibfnamefont {L.}~\bibnamefont
  {Yuan}}, \bibinfo {author} {\bibfnamefont {Y.}~\bibnamefont {Liu}}, \bibinfo
  {author} {\bibfnamefont {J.}~\bibnamefont {Yao}},\ and\ \bibinfo {author}
  {\bibfnamefont {Y.}~\bibnamefont {Cheng}},\ }\bibfield  {title} {\bibinfo
  {title} {Recent {Advances} in {Air} {Lasing}: {A} {Perspective} from
  {Quantum} {Coherence}},\ }\href {https://doi.org/10.1002/qute.201900080}
  {\bibfield  {journal} {\bibinfo  {journal} {Adv Quantum Tech}\ }\textbf
  {\bibinfo {volume} {2}},\ \bibinfo {pages} {1900080} (\bibinfo {year}
  {2019})}\BibitemShut {NoStop}%
\bibitem [{\citenamefont {Ando}\ \emph {et~al.}(2019)\citenamefont {Ando},
  \citenamefont {Lötstedt}, \citenamefont {Iwasaki}, \citenamefont {Li},
  \citenamefont {Fu}, \citenamefont {Wang}, \citenamefont {Xu},\ and\
  \citenamefont {Yamanouchi}}]{ando_rotational_2019}%
  \BibitemOpen
  \bibfield  {author} {\bibinfo {author} {\bibfnamefont {T.}~\bibnamefont
  {Ando}}, \bibinfo {author} {\bibfnamefont {E.}~\bibnamefont {Lötstedt}},
  \bibinfo {author} {\bibfnamefont {A.}~\bibnamefont {Iwasaki}}, \bibinfo
  {author} {\bibfnamefont {H.}~\bibnamefont {Li}}, \bibinfo {author}
  {\bibfnamefont {Y.}~\bibnamefont {Fu}}, \bibinfo {author} {\bibfnamefont
  {S.}~\bibnamefont {Wang}}, \bibinfo {author} {\bibfnamefont {H.}~\bibnamefont
  {Xu}},\ and\ \bibinfo {author} {\bibfnamefont {K.}~\bibnamefont
  {Yamanouchi}},\ }\bibfield  {title} {\bibinfo {title} {Rotational,
  {Vibrational}, and {Electronic} {Modulations} in n$_2^+$ {Lasing} at 391 nm:
  {Evidence} of {Coherent} {B$^2\Sigma_U^+$}$-${X$^2\Sigma_g^+$}$-${A$^2\Pi_u$}
  {Coupling}},\ }\href {https://doi.org/10.1103/PhysRevLett.123.203201}
  {\bibfield  {journal} {\bibinfo  {journal} {Phys. Rev. Lett.}\ }\textbf
  {\bibinfo {volume} {123}},\ \bibinfo {pages} {203201} (\bibinfo {year}
  {2019})}\BibitemShut {NoStop}%
\bibitem [{\citenamefont {Danylo}\ \emph {et~al.}(2019)\citenamefont {Danylo},
  \citenamefont {Zhang}, \citenamefont {Fan}, \citenamefont {Zhou},
  \citenamefont {Lu}, \citenamefont {Zhou}, \citenamefont {Liang},
  \citenamefont {Zhuang}, \citenamefont {Houard}, \citenamefont {Mysyrowicz},
  \citenamefont {Oliva},\ and\ \citenamefont {Liu}}]{danylo_formation_2019}%
  \BibitemOpen
  \bibfield  {author} {\bibinfo {author} {\bibfnamefont {R.}~\bibnamefont
  {Danylo}}, \bibinfo {author} {\bibfnamefont {X.}~\bibnamefont {Zhang}},
  \bibinfo {author} {\bibfnamefont {Z.}~\bibnamefont {Fan}}, \bibinfo {author}
  {\bibfnamefont {D.}~\bibnamefont {Zhou}}, \bibinfo {author} {\bibfnamefont
  {Q.}~\bibnamefont {Lu}}, \bibinfo {author} {\bibfnamefont {B.}~\bibnamefont
  {Zhou}}, \bibinfo {author} {\bibfnamefont {Q.}~\bibnamefont {Liang}},
  \bibinfo {author} {\bibfnamefont {S.}~\bibnamefont {Zhuang}}, \bibinfo
  {author} {\bibfnamefont {A.}~\bibnamefont {Houard}}, \bibinfo {author}
  {\bibfnamefont {A.}~\bibnamefont {Mysyrowicz}}, \bibinfo {author}
  {\bibfnamefont {E.}~\bibnamefont {Oliva}},\ and\ \bibinfo {author}
  {\bibfnamefont {Y.}~\bibnamefont {Liu}},\ }\bibfield  {title} {\bibinfo
  {title} {Formation {Dynamics} of {Excited} {Neutral} {Nitrogen} {Molecules}
  inside {Femtosecond} {Laser} {Filaments}},\ }\href
  {https://doi.org/10.1103/PhysRevLett.123.243203} {\bibfield  {journal}
  {\bibinfo  {journal} {Phys. Rev. Lett.}\ }\textbf {\bibinfo {volume} {123}},\
  \bibinfo {pages} {243203} (\bibinfo {year} {2019})}\BibitemShut {NoStop}%
\bibitem [{\citenamefont {Li}\ \emph {et~al.}(2019)\citenamefont {Li},
  \citenamefont {Hou}, \citenamefont {Zang}, \citenamefont {Fu}, \citenamefont
  {Lötstedt}, \citenamefont {Ando}, \citenamefont {Iwasaki}, \citenamefont
  {Yamanouchi},\ and\ \citenamefont {Xu}}]{li_significant_2019}%
  \BibitemOpen
  \bibfield  {author} {\bibinfo {author} {\bibfnamefont {H.}~\bibnamefont
  {Li}}, \bibinfo {author} {\bibfnamefont {M.}~\bibnamefont {Hou}}, \bibinfo
  {author} {\bibfnamefont {H.}~\bibnamefont {Zang}}, \bibinfo {author}
  {\bibfnamefont {Y.}~\bibnamefont {Fu}}, \bibinfo {author} {\bibfnamefont
  {E.}~\bibnamefont {Lötstedt}}, \bibinfo {author} {\bibfnamefont
  {T.}~\bibnamefont {Ando}}, \bibinfo {author} {\bibfnamefont {A.}~\bibnamefont
  {Iwasaki}}, \bibinfo {author} {\bibfnamefont {K.}~\bibnamefont
  {Yamanouchi}},\ and\ \bibinfo {author} {\bibfnamefont {H.}~\bibnamefont
  {Xu}},\ }\bibfield  {title} {\bibinfo {title} {Significant {Enhancement} of
  {N} 2 + {Lasing} by {Polarization}-{Modulated} {Ultrashort} {Laser}
  {Pulses}},\ }\href {https://doi.org/10.1103/PhysRevLett.122.013202}
  {\bibfield  {journal} {\bibinfo  {journal} {Phys. Rev. Lett.}\ }\textbf
  {\bibinfo {volume} {122}},\ \bibinfo {pages} {013202} (\bibinfo {year}
  {2019})}\BibitemShut {NoStop}%
\bibitem [{\citenamefont {Li}\ \emph {et~al.}(2020)\citenamefont {Li},
  \citenamefont {Lötstedt}, \citenamefont {Li}, \citenamefont {Zhou},
  \citenamefont {Dong}, \citenamefont {Deng}, \citenamefont {Lu}, \citenamefont
  {Ando}, \citenamefont {Iwasaki}, \citenamefont {Fu}, \citenamefont {Wang},
  \citenamefont {Wu}, \citenamefont {Yamanouchi},\ and\ \citenamefont
  {Xu}}]{li_giant_2020}%
  \BibitemOpen
  \bibfield  {author} {\bibinfo {author} {\bibfnamefont {H.}~\bibnamefont
  {Li}}, \bibinfo {author} {\bibfnamefont {E.}~\bibnamefont {Lötstedt}},
  \bibinfo {author} {\bibfnamefont {H.}~\bibnamefont {Li}}, \bibinfo {author}
  {\bibfnamefont {Y.}~\bibnamefont {Zhou}}, \bibinfo {author} {\bibfnamefont
  {N.}~\bibnamefont {Dong}}, \bibinfo {author} {\bibfnamefont {L.}~\bibnamefont
  {Deng}}, \bibinfo {author} {\bibfnamefont {P.}~\bibnamefont {Lu}}, \bibinfo
  {author} {\bibfnamefont {T.}~\bibnamefont {Ando}}, \bibinfo {author}
  {\bibfnamefont {A.}~\bibnamefont {Iwasaki}}, \bibinfo {author} {\bibfnamefont
  {Y.}~\bibnamefont {Fu}}, \bibinfo {author} {\bibfnamefont {S.}~\bibnamefont
  {Wang}}, \bibinfo {author} {\bibfnamefont {J.}~\bibnamefont {Wu}}, \bibinfo
  {author} {\bibfnamefont {K.}~\bibnamefont {Yamanouchi}},\ and\ \bibinfo
  {author} {\bibfnamefont {H.}~\bibnamefont {Xu}},\ }\bibfield  {title}
  {\bibinfo {title} {Giant {Enhancement} of {Air} {Lasing} by {Complete}
  {Population} {Inversion} in {N} 2 +},\ }\href
  {https://doi.org/10.1103/PhysRevLett.125.053201} {\bibfield  {journal}
  {\bibinfo  {journal} {Phys. Rev. Lett.}\ }\textbf {\bibinfo {volume} {125}},\
  \bibinfo {pages} {053201} (\bibinfo {year} {2020})}\BibitemShut {NoStop}%
\bibitem [{\citenamefont {Richter}\ \emph {et~al.}(2020)\citenamefont
  {Richter}, \citenamefont {Lytova}, \citenamefont {Morales}, \citenamefont
  {Haessler}, \citenamefont {Smirnova}, \citenamefont {Spanner},\ and\
  \citenamefont {Ivanov}}]{richter_rotational_2020}%
  \BibitemOpen
  \bibfield  {author} {\bibinfo {author} {\bibfnamefont {M.}~\bibnamefont
  {Richter}}, \bibinfo {author} {\bibfnamefont {M.}~\bibnamefont {Lytova}},
  \bibinfo {author} {\bibfnamefont {F.}~\bibnamefont {Morales}}, \bibinfo
  {author} {\bibfnamefont {S.}~\bibnamefont {Haessler}}, \bibinfo {author}
  {\bibfnamefont {O.}~\bibnamefont {Smirnova}}, \bibinfo {author}
  {\bibfnamefont {M.}~\bibnamefont {Spanner}},\ and\ \bibinfo {author}
  {\bibfnamefont {M.}~\bibnamefont {Ivanov}},\ }\bibfield  {title} {\bibinfo
  {title} {Rotational quantum beat lasing without inversion},\ }\href
  {https://doi.org/10.1364/OPTICA.390665} {\bibfield  {journal} {\bibinfo
  {journal} {Optica}\ }\textbf {\bibinfo {volume} {7}},\ \bibinfo {pages} {586}
  (\bibinfo {year} {2020})}\BibitemShut {NoStop}%
\bibitem [{\citenamefont {Zhang}\ \emph {et~al.}(2021)\citenamefont {Zhang},
  \citenamefont {Lu}, \citenamefont {Zhang}, \citenamefont {Fan}, \citenamefont
  {Zhou}, \citenamefont {Liang}, \citenamefont {Yuan}, \citenamefont {Zhuang},
  \citenamefont {Dorfman},\ and\ \citenamefont {Liu}}]{zhang_coherent_2021}%
  \BibitemOpen
  \bibfield  {author} {\bibinfo {author} {\bibfnamefont {X.}~\bibnamefont
  {Zhang}}, \bibinfo {author} {\bibfnamefont {Q.}~\bibnamefont {Lu}}, \bibinfo
  {author} {\bibfnamefont {Z.}~\bibnamefont {Zhang}}, \bibinfo {author}
  {\bibfnamefont {Z.}~\bibnamefont {Fan}}, \bibinfo {author} {\bibfnamefont
  {D.}~\bibnamefont {Zhou}}, \bibinfo {author} {\bibfnamefont {Q.}~\bibnamefont
  {Liang}}, \bibinfo {author} {\bibfnamefont {L.}~\bibnamefont {Yuan}},
  \bibinfo {author} {\bibfnamefont {S.}~\bibnamefont {Zhuang}}, \bibinfo
  {author} {\bibfnamefont {K.}~\bibnamefont {Dorfman}},\ and\ \bibinfo {author}
  {\bibfnamefont {Y.}~\bibnamefont {Liu}},\ }\bibfield  {title} {\bibinfo
  {title} {Coherent control of the multiple wavelength lasing of {N}$_2^ +$:
  coherence transfer and beyond},\ }\href
  {https://doi.org/10.1364/OPTICA.417804} {\bibfield  {journal} {\bibinfo
  {journal} {Optica}\ }\textbf {\bibinfo {volume} {8}},\ \bibinfo {pages} {668}
  (\bibinfo {year} {2021})}\BibitemShut {NoStop}%
\bibitem [{\citenamefont {Kleine}\ \emph {et~al.}(2022)\citenamefont {Kleine},
  \citenamefont {Winghart}, \citenamefont {Zhang}, \citenamefont {Richter},
  \citenamefont {Ekimova}, \citenamefont {Eckert}, \citenamefont {Vrakking},
  \citenamefont {Nibbering}, \citenamefont {Rouzée},\ and\ \citenamefont
  {Grant}}]{kleine_electronic_2022}%
  \BibitemOpen
  \bibfield  {author} {\bibinfo {author} {\bibfnamefont {C.}~\bibnamefont
  {Kleine}}, \bibinfo {author} {\bibfnamefont {M.-O.}\ \bibnamefont
  {Winghart}}, \bibinfo {author} {\bibfnamefont {Z.-Y.}\ \bibnamefont {Zhang}},
  \bibinfo {author} {\bibfnamefont {M.}~\bibnamefont {Richter}}, \bibinfo
  {author} {\bibfnamefont {M.}~\bibnamefont {Ekimova}}, \bibinfo {author}
  {\bibfnamefont {S.}~\bibnamefont {Eckert}}, \bibinfo {author} {\bibfnamefont
  {M.~J.}\ \bibnamefont {Vrakking}}, \bibinfo {author} {\bibfnamefont {E.~T.}\
  \bibnamefont {Nibbering}}, \bibinfo {author} {\bibfnamefont {A.}~\bibnamefont
  {Rouzée}},\ and\ \bibinfo {author} {\bibfnamefont {E.~R.}\ \bibnamefont
  {Grant}},\ }\bibfield  {title} {\bibinfo {title} {Electronic {State}
  {Population} {Dynamics} upon {Ultrafast} {Strong} {Field} {Ionization} and
  {Fragmentation} of {Molecular} {Nitrogen}},\ }\href
  {https://doi.org/10.1103/PhysRevLett.129.123002} {\bibfield  {journal}
  {\bibinfo  {journal} {Phys. Rev. Lett.}\ }\textbf {\bibinfo {volume} {129}},\
  \bibinfo {pages} {123002} (\bibinfo {year} {2022})}\BibitemShut {NoStop}%
\bibitem [{\citenamefont {Zhuang}\ \emph {et~al.}(2022)\citenamefont {Zhuang},
  \citenamefont {Zhang}, \citenamefont {Lu},\ and\ \citenamefont
  {Liu}}]{zhuang_optical_2022}%
  \BibitemOpen
  \bibfield  {author} {\bibinfo {author} {\bibfnamefont {C.}~\bibnamefont
  {Zhuang}}, \bibinfo {author} {\bibfnamefont {X.}~\bibnamefont {Zhang}},
  \bibinfo {author} {\bibfnamefont {Q.}~\bibnamefont {Lu}},\ and\ \bibinfo
  {author} {\bibfnamefont {Y.}~\bibnamefont {Liu}},\ }\bibfield  {title}
  {\bibinfo {title} {Optical amplification and gain dynamics of cavity-free
  lasing of argon pumped by ultraviolet femtosecond pulses},\ }\href
  {https://doi.org/10.1364/OE.455743} {\bibfield  {journal} {\bibinfo
  {journal} {Opt. Express}\ }\textbf {\bibinfo {volume} {30}},\ \bibinfo
  {pages} {17156} (\bibinfo {year} {2022})}\BibitemShut {NoStop}%
\bibitem [{\citenamefont {Dogariu}\ and\ \citenamefont
  {Miles}(2015)}]{dogariu_remote_2015}%
  \BibitemOpen
  \bibfield  {author} {\bibinfo {author} {\bibfnamefont {A.}~\bibnamefont
  {Dogariu}}\ and\ \bibinfo {author} {\bibfnamefont {R.~B.}\ \bibnamefont
  {Miles}},\ }\bibfield  {title} {\bibinfo {title} {Remote
  {Backward}-{Propagating} {Lasing} of {Nitrogen} and {Oxygen} in {Air}},\ }in\
  \href {https://doi.org/10.1364/CLEO_SI.2015.SM1N.1} {\emph {\bibinfo
  {booktitle} {{CLEO}: 2015}}}\ (\bibinfo  {publisher} {OSA},\ \bibinfo
  {address} {San Jose, California},\ \bibinfo {year} {2015})\ p.\ \bibinfo
  {pages} {SM1N.1}\BibitemShut {NoStop}%
\bibitem [{\citenamefont {Dogariu}\ and\ \citenamefont
  {Miles}(2018)}]{dogariu_backwards_2018}%
  \BibitemOpen
  \bibfield  {author} {\bibinfo {author} {\bibfnamefont {A.}~\bibnamefont
  {Dogariu}}\ and\ \bibinfo {author} {\bibfnamefont {R.~B.}\ \bibnamefont
  {Miles}},\ }\bibfield  {title} {\bibinfo {title} {Backwards {Lasing} in
  {Atmospheric} {Air} from {Argon}},\ }in\ \href
  {https://doi.org/10.1364/CLEO_AT.2018.JTh5B.8} {\emph {\bibinfo {booktitle}
  {Conference on {Lasers} and {Electro}-{Optics}}}}\ (\bibinfo  {publisher}
  {OSA},\ \bibinfo {address} {San Jose, California},\ \bibinfo {year} {2018})\
  p.\ \bibinfo {pages} {JTh5B.8}\BibitemShut {NoStop}%
\bibitem [{\citenamefont {Grib}\ \emph {et~al.}(2021)\citenamefont {Grib},
  \citenamefont {Stauffer}, \citenamefont {Roy},\ and\ \citenamefont
  {Schumaker}}]{grib_resonance-enhanced_2021}%
  \BibitemOpen
  \bibfield  {author} {\bibinfo {author} {\bibfnamefont {S.~W.}\ \bibnamefont
  {Grib}}, \bibinfo {author} {\bibfnamefont {H.~U.}\ \bibnamefont {Stauffer}},
  \bibinfo {author} {\bibfnamefont {S.}~\bibnamefont {Roy}},\ and\ \bibinfo
  {author} {\bibfnamefont {S.~A.}\ \bibnamefont {Schumaker}},\ }\bibfield
  {title} {\bibinfo {title} {Resonance-enhanced, rare-gas-assisted
  femtosecond-laser electronic-excitation tagging in argon/nitrogen mixtures},\
  }\href {https://doi.org/10.1364/AO.419125} {\bibfield  {journal} {\bibinfo
  {journal} {Appl. Opt.}\ }\textbf {\bibinfo {volume} {60}},\ \bibinfo {pages}
  {C32} (\bibinfo {year} {2021})}\BibitemShut {NoStop}%
\bibitem [{\citenamefont {Michael}\ \emph {et~al.}(2011)\citenamefont
  {Michael}, \citenamefont {Edwards}, \citenamefont {Dogariu},\ and\
  \citenamefont {Miles}}]{michael_femtosecond_2011}%
  \BibitemOpen
  \bibfield  {author} {\bibinfo {author} {\bibfnamefont {J.~B.}\ \bibnamefont
  {Michael}}, \bibinfo {author} {\bibfnamefont {M.~R.}\ \bibnamefont
  {Edwards}}, \bibinfo {author} {\bibfnamefont {A.}~\bibnamefont {Dogariu}},\
  and\ \bibinfo {author} {\bibfnamefont {R.~B.}\ \bibnamefont {Miles}},\
  }\bibfield  {title} {\bibinfo {title} {Femtosecond laser electronic
  excitation tagging for quantitative velocity imaging in air},\ }\href
  {https://doi.org/10.1364/AO.50.005158} {\bibfield  {journal} {\bibinfo
  {journal} {Appl. Opt.}\ }\textbf {\bibinfo {volume} {50}},\ \bibinfo {pages}
  {5158} (\bibinfo {year} {2011})}\BibitemShut {NoStop}%
\bibitem [{\citenamefont {Dicke}(1954)}]{dicke_coherence_1954}%
  \BibitemOpen
  \bibfield  {author} {\bibinfo {author} {\bibfnamefont {R.~H.}\ \bibnamefont
  {Dicke}},\ }\bibfield  {title} {\bibinfo {title} {Coherence in {Spontaneous}
  {Radiation} {Processes}},\ }\href {https://doi.org/10.1103/PhysRev.93.99}
  {\bibfield  {journal} {\bibinfo  {journal} {Phys. Rev.}\ }\textbf {\bibinfo
  {volume} {93}},\ \bibinfo {pages} {99} (\bibinfo {year} {1954})}\BibitemShut
  {NoStop}%
\bibitem [{\citenamefont {Maki}\ \emph {et~al.}(1989)\citenamefont {Maki},
  \citenamefont {Malcuit}, \citenamefont {Raymer}, \citenamefont {Boyd},\ and\
  \citenamefont {Drummond}}]{maki_influence_1989}%
  \BibitemOpen
  \bibfield  {author} {\bibinfo {author} {\bibfnamefont {J.~J.}\ \bibnamefont
  {Maki}}, \bibinfo {author} {\bibfnamefont {M.~S.}\ \bibnamefont {Malcuit}},
  \bibinfo {author} {\bibfnamefont {M.~G.}\ \bibnamefont {Raymer}}, \bibinfo
  {author} {\bibfnamefont {R.~W.}\ \bibnamefont {Boyd}},\ and\ \bibinfo
  {author} {\bibfnamefont {P.~D.}\ \bibnamefont {Drummond}},\ }\bibfield
  {title} {\bibinfo {title} {Influence of collisional dephasing processes on
  superfluorescence},\ }\href {https://doi.org/10.1103/PhysRevA.40.5135}
  {\bibfield  {journal} {\bibinfo  {journal} {Phys. Rev. A}\ }\textbf {\bibinfo
  {volume} {40}},\ \bibinfo {pages} {5135} (\bibinfo {year}
  {1989})}\BibitemShut {NoStop}%
\bibitem [{\citenamefont {Cederbaum}\ \emph {et~al.}(1997)\citenamefont
  {Cederbaum}, \citenamefont {Zobeley},\ and\ \citenamefont
  {Tarantelli}}]{cederbaum_giant_1997}%
  \BibitemOpen
  \bibfield  {author} {\bibinfo {author} {\bibfnamefont {L.~S.}\ \bibnamefont
  {Cederbaum}}, \bibinfo {author} {\bibfnamefont {J.}~\bibnamefont {Zobeley}},\
  and\ \bibinfo {author} {\bibfnamefont {F.}~\bibnamefont {Tarantelli}},\
  }\bibfield  {title} {\bibinfo {title} {Giant {Intermolecular} {Decay} and
  {Fragmentation} of {Clusters}},\ }\href
  {https://doi.org/10.1103/PhysRevLett.79.4778} {\bibfield  {journal} {\bibinfo
   {journal} {Phys. Rev. Lett.}\ }\textbf {\bibinfo {volume} {79}},\ \bibinfo
  {pages} {4778} (\bibinfo {year} {1997})}\BibitemShut {NoStop}%
\bibitem [{\citenamefont {Santra}\ \emph {et~al.}(2000)\citenamefont {Santra},
  \citenamefont {Zobeley}, \citenamefont {Cederbaum},\ and\ \citenamefont
  {Moiseyev}}]{santra_interatomic_2000}%
  \BibitemOpen
  \bibfield  {author} {\bibinfo {author} {\bibfnamefont {R.}~\bibnamefont
  {Santra}}, \bibinfo {author} {\bibfnamefont {J.}~\bibnamefont {Zobeley}},
  \bibinfo {author} {\bibfnamefont {L.~S.}\ \bibnamefont {Cederbaum}},\ and\
  \bibinfo {author} {\bibfnamefont {N.}~\bibnamefont {Moiseyev}},\ }\bibfield
  {title} {\bibinfo {title} {Interatomic {Coulombic} {Decay} in van der {Waals}
  {Clusters} and {Impact} of {Nuclear} {Motion}},\ }\href
  {https://doi.org/10.1103/PhysRevLett.85.4490} {\bibfield  {journal} {\bibinfo
   {journal} {Phys. Rev. Lett.}\ }\textbf {\bibinfo {volume} {85}},\ \bibinfo
  {pages} {4490} (\bibinfo {year} {2000})}\BibitemShut {NoStop}%
\bibitem [{\citenamefont {Averbukh}\ \emph {et~al.}(2004)\citenamefont
  {Averbukh}, \citenamefont {Müller},\ and\ \citenamefont
  {Cederbaum}}]{averbukh_mechanism_2004}%
  \BibitemOpen
  \bibfield  {author} {\bibinfo {author} {\bibfnamefont {V.}~\bibnamefont
  {Averbukh}}, \bibinfo {author} {\bibfnamefont {I.~B.}\ \bibnamefont
  {Müller}},\ and\ \bibinfo {author} {\bibfnamefont {L.~S.}\ \bibnamefont
  {Cederbaum}},\ }\bibfield  {title} {\bibinfo {title} {Mechanism of
  {Interatomic} {Coulombic} {Decay} in {Clusters}},\ }\href
  {https://doi.org/10.1103/PhysRevLett.93.263002} {\bibfield  {journal}
  {\bibinfo  {journal} {Phys. Rev. Lett.}\ }\textbf {\bibinfo {volume} {93}},\
  \bibinfo {pages} {263002} (\bibinfo {year} {2004})}\BibitemShut {NoStop}%
\bibitem [{\citenamefont {Jahnke}\ \emph {et~al.}(2004)\citenamefont {Jahnke},
  \citenamefont {Czasch}, \citenamefont {Schöffler}, \citenamefont
  {Schössler}, \citenamefont {Knapp}, \citenamefont {Käsz}, \citenamefont
  {Titze}, \citenamefont {Wimmer}, \citenamefont {Kreidi}, \citenamefont
  {Grisenti}, \citenamefont {Staudte}, \citenamefont {Jagutzki}, \citenamefont
  {Hergenhahn}, \citenamefont {Schmidt-Böcking},\ and\ \citenamefont
  {Dörner}}]{jahnke_experimental_2004}%
  \BibitemOpen
  \bibfield  {author} {\bibinfo {author} {\bibfnamefont {T.}~\bibnamefont
  {Jahnke}}, \bibinfo {author} {\bibfnamefont {A.}~\bibnamefont {Czasch}},
  \bibinfo {author} {\bibfnamefont {M.~S.}\ \bibnamefont {Schöffler}},
  \bibinfo {author} {\bibfnamefont {S.}~\bibnamefont {Schössler}}, \bibinfo
  {author} {\bibfnamefont {A.}~\bibnamefont {Knapp}}, \bibinfo {author}
  {\bibfnamefont {M.}~\bibnamefont {Käsz}}, \bibinfo {author} {\bibfnamefont
  {J.}~\bibnamefont {Titze}}, \bibinfo {author} {\bibfnamefont
  {C.}~\bibnamefont {Wimmer}}, \bibinfo {author} {\bibfnamefont
  {K.}~\bibnamefont {Kreidi}}, \bibinfo {author} {\bibfnamefont {R.~E.}\
  \bibnamefont {Grisenti}}, \bibinfo {author} {\bibfnamefont {A.}~\bibnamefont
  {Staudte}}, \bibinfo {author} {\bibfnamefont {O.}~\bibnamefont {Jagutzki}},
  \bibinfo {author} {\bibfnamefont {U.}~\bibnamefont {Hergenhahn}}, \bibinfo
  {author} {\bibfnamefont {H.}~\bibnamefont {Schmidt-Böcking}},\ and\ \bibinfo
  {author} {\bibfnamefont {R.}~\bibnamefont {Dörner}},\ }\bibfield  {title}
  {\bibinfo {title} {Experimental {Observation} of {Interatomic} {Coulombic}
  {Decay} in {Neon} {Dimers}},\ }\href
  {https://doi.org/10.1103/PhysRevLett.93.163401} {\bibfield  {journal}
  {\bibinfo  {journal} {Phys. Rev. Lett.}\ }\textbf {\bibinfo {volume} {93}},\
  \bibinfo {pages} {163401} (\bibinfo {year} {2004})}\BibitemShut {NoStop}%
\bibitem [{\citenamefont {Kitano}\ and\ \citenamefont
  {Maeda}(2023)}]{kitano_cascade_2023}%
  \BibitemOpen
  \bibfield  {author} {\bibinfo {author} {\bibfnamefont {K.}~\bibnamefont
  {Kitano}}\ and\ \bibinfo {author} {\bibfnamefont {H.}~\bibnamefont {Maeda}},\
  }\bibfield  {title} {\bibinfo {title} {Cascade and yoked superfluorescence
  detected by sum frequency generation spectroscopy},\ }\href
  {https://doi.org/10.1364/OL.473200} {\bibfield  {journal} {\bibinfo
  {journal} {Opt. Lett.}\ }\textbf {\bibinfo {volume} {48}},\ \bibinfo {pages}
  {69} (\bibinfo {year} {2023})}\BibitemShut {NoStop}%
\bibitem [{\citenamefont {Brownell}\ \emph {et~al.}(1995)\citenamefont
  {Brownell}, \citenamefont {Lu},\ and\ \citenamefont
  {Hartmann}}]{brownell_yoked_1995}%
  \BibitemOpen
  \bibfield  {author} {\bibinfo {author} {\bibfnamefont {J.~H.}\ \bibnamefont
  {Brownell}}, \bibinfo {author} {\bibfnamefont {X.}~\bibnamefont {Lu}},\ and\
  \bibinfo {author} {\bibfnamefont {S.~R.}\ \bibnamefont {Hartmann}},\
  }\bibfield  {title} {\bibinfo {title} {Yoked {Superfluorescence}},\ }\href
  {https://doi.org/10.1103/PhysRevLett.75.3265} {\bibfield  {journal} {\bibinfo
   {journal} {Phys. Rev. Lett.}\ }\textbf {\bibinfo {volume} {75}},\ \bibinfo
  {pages} {3265} (\bibinfo {year} {1995})}\BibitemShut {NoStop}%
\bibitem [{\citenamefont {Verdeyen}(1995)}]{verdeyen_laser_1995}%
  \BibitemOpen
  \bibfield  {author} {\bibinfo {author} {\bibfnamefont {J.~T.}\ \bibnamefont
  {Verdeyen}},\ }\href@noop {} {\emph {\bibinfo {title} {Laser electronics}}},\
  \bibinfo {edition} {3rd}\ ed.,\ Prentice {Hall} series in solid state
  physical electronics\ (\bibinfo  {publisher} {Prentice Hall},\ \bibinfo
  {address} {Englewood Cliffs, N.J},\ \bibinfo {year} {1995})\BibitemShut
  {NoStop}%
\bibitem [{\citenamefont {Kunze}\ \emph {et~al.}(2022)\citenamefont {Kunze},
  \citenamefont {Groll}, \citenamefont {Besser},\ and\ \citenamefont
  {Thöming}}]{kunze_molecular_2022}%
  \BibitemOpen
  \bibfield  {author} {\bibinfo {author} {\bibfnamefont {S.}~\bibnamefont
  {Kunze}}, \bibinfo {author} {\bibfnamefont {R.}~\bibnamefont {Groll}},
  \bibinfo {author} {\bibfnamefont {B.}~\bibnamefont {Besser}},\ and\ \bibinfo
  {author} {\bibfnamefont {J.}~\bibnamefont {Thöming}},\ }\bibfield  {title}
  {\bibinfo {title} {Molecular diameters of rarefied gases},\ }\href
  {https://doi.org/10.1038/s41598-022-05871-y} {\bibfield  {journal} {\bibinfo
  {journal} {Sci Rep}\ }\textbf {\bibinfo {volume} {12}},\ \bibinfo {pages}
  {2057} (\bibinfo {year} {2022})}\BibitemShut {NoStop}%
\bibitem [{\citenamefont {Grynberg}(1979)}]{grynberg_three-photon_1979}%
  \BibitemOpen
  \bibfield  {author} {\bibinfo {author} {\bibfnamefont {G.}~\bibnamefont
  {Grynberg}},\ }\bibfield  {title} {\bibinfo {title} {Three-photon absorption
  : selection rules and line intensities},\ }\href
  {https://doi.org/10.1051/jphys:019790040010096500} {\bibfield  {journal}
  {\bibinfo  {journal} {J. Phys. France}\ }\textbf {\bibinfo {volume} {40}},\
  \bibinfo {pages} {965} (\bibinfo {year} {1979})}\BibitemShut {NoStop}%
\bibitem [{\citenamefont {Jackson}\ and\ \citenamefont
  {Wynne}(1982)}]{jackson_interference_1982}%
  \BibitemOpen
  \bibfield  {author} {\bibinfo {author} {\bibfnamefont {D.~J.}\ \bibnamefont
  {Jackson}}\ and\ \bibinfo {author} {\bibfnamefont {J.~J.}\ \bibnamefont
  {Wynne}},\ }\bibfield  {title} {\bibinfo {title} {Interference {Effects}
  between {Different} {Optical} {Harmonics}},\ }\href@noop {} {\bibfield
  {journal} {\bibinfo  {journal} {Phys. Rev. Lett.}\ }\textbf {\bibinfo
  {volume} {49}},\ \bibinfo {pages} {543} (\bibinfo {year} {1982})}\BibitemShut
  {NoStop}%
\bibitem [{\citenamefont {Payne}\ \emph {et~al.}(1980)\citenamefont {Payne},
  \citenamefont {Garett},\ and\ \citenamefont {Baker}}]{payne_effects_1980}%
  \BibitemOpen
  \bibfield  {author} {\bibinfo {author} {\bibfnamefont {M.~G.}\ \bibnamefont
  {Payne}}, \bibinfo {author} {\bibfnamefont {W.~R.}\ \bibnamefont {Garett}},\
  and\ \bibinfo {author} {\bibfnamefont {H.~C.}\ \bibnamefont {Baker}},\
  }\bibfield  {title} {\bibinfo {title} {Effects of collective emission on
  multiphoton excitation and ionization near a three-photon resonance},\ }\href
  {https://doi.org/10.1016/0009-2614(80)80557-4} {\bibfield  {journal}
  {\bibinfo  {journal} {Chemical Physics Letters}\ }\textbf {\bibinfo {volume}
  {75}},\ \bibinfo {pages} {468} (\bibinfo {year} {1980})}\BibitemShut
  {NoStop}%
\bibitem [{\citenamefont {Gordon}\ \emph {et~al.}(2022)\citenamefont {Gordon},
  \citenamefont {Rothman}, \citenamefont {Hargreaves}, \citenamefont {Hashemi},
  \citenamefont {Karlovets}, \citenamefont {Skinner}, \citenamefont {Conway},
  \citenamefont {Hill}, \citenamefont {Kochanov}, \citenamefont {Tan},
  \citenamefont {Wcisło}, \citenamefont {Finenko}, \citenamefont {Nelson},
  \citenamefont {Bernath}, \citenamefont {Birk}, \citenamefont {Boudon},
  \citenamefont {Campargue}, \citenamefont {Chance}, \citenamefont {Coustenis},
  \citenamefont {Drouin}, \citenamefont {Flaud}, \citenamefont {Gamache},
  \citenamefont {Hodges}, \citenamefont {Jacquemart}, \citenamefont {Mlawer},
  \citenamefont {Nikitin}, \citenamefont {Perevalov}, \citenamefont {Rotger},
  \citenamefont {Tennyson}, \citenamefont {Toon}, \citenamefont {Tran},
  \citenamefont {Tyuterev}, \citenamefont {Adkins}, \citenamefont {Baker},
  \citenamefont {Barbe}, \citenamefont {Canè}, \citenamefont {Császár},
  \citenamefont {Dudaryonok}, \citenamefont {Egorov}, \citenamefont {Fleisher},
  \citenamefont {Fleurbaey}, \citenamefont {Foltynowicz}, \citenamefont
  {Furtenbacher}, \citenamefont {Harrison}, \citenamefont {Hartmann},
  \citenamefont {Horneman}, \citenamefont {Huang}, \citenamefont {Karman},
  \citenamefont {Karns}, \citenamefont {Kassi}, \citenamefont {Kleiner},
  \citenamefont {Kofman}, \citenamefont {Kwabia–Tchana}, \citenamefont
  {Lavrentieva}, \citenamefont {Lee}, \citenamefont {Long}, \citenamefont
  {Lukashevskaya}, \citenamefont {Lyulin}, \citenamefont {Makhnev},
  \citenamefont {Matt}, \citenamefont {Massie}, \citenamefont {Melosso},
  \citenamefont {Mikhailenko}, \citenamefont {Mondelain}, \citenamefont
  {Müller}, \citenamefont {Naumenko}, \citenamefont {Perrin}, \citenamefont
  {Polyansky}, \citenamefont {Raddaoui}, \citenamefont {Raston}, \citenamefont
  {Reed}, \citenamefont {Rey}, \citenamefont {Richard}, \citenamefont
  {Tóbiás}, \citenamefont {Sadiek}, \citenamefont {Schwenke}, \citenamefont
  {Starikova}, \citenamefont {Sung}, \citenamefont {Tamassia}, \citenamefont
  {Tashkun}, \citenamefont {Vander~Auwera}, \citenamefont {Vasilenko},
  \citenamefont {Vigasin}, \citenamefont {Villanueva}, \citenamefont {Vispoel},
  \citenamefont {Wagner}, \citenamefont {Yachmenev},\ and\ \citenamefont
  {Yurchenko}}]{gordon_hitran2020_2022}%
  \BibitemOpen
  \bibfield  {author} {\bibinfo {author} {\bibfnamefont {I.}~\bibnamefont
  {Gordon}}, \bibinfo {author} {\bibfnamefont {L.}~\bibnamefont {Rothman}},
  \bibinfo {author} {\bibfnamefont {R.}~\bibnamefont {Hargreaves}}, \bibinfo
  {author} {\bibfnamefont {R.}~\bibnamefont {Hashemi}}, \bibinfo {author}
  {\bibfnamefont {E.}~\bibnamefont {Karlovets}}, \bibinfo {author}
  {\bibfnamefont {F.}~\bibnamefont {Skinner}}, \bibinfo {author} {\bibfnamefont
  {E.}~\bibnamefont {Conway}}, \bibinfo {author} {\bibfnamefont
  {C.}~\bibnamefont {Hill}}, \bibinfo {author} {\bibfnamefont {R.}~\bibnamefont
  {Kochanov}}, \bibinfo {author} {\bibfnamefont {Y.}~\bibnamefont {Tan}},
  \bibinfo {author} {\bibfnamefont {P.}~\bibnamefont {Wcisło}}, \bibinfo
  {author} {\bibfnamefont {A.}~\bibnamefont {Finenko}}, \bibinfo {author}
  {\bibfnamefont {K.}~\bibnamefont {Nelson}}, \bibinfo {author} {\bibfnamefont
  {P.}~\bibnamefont {Bernath}}, \bibinfo {author} {\bibfnamefont
  {M.}~\bibnamefont {Birk}}, \bibinfo {author} {\bibfnamefont {V.}~\bibnamefont
  {Boudon}}, \bibinfo {author} {\bibfnamefont {A.}~\bibnamefont {Campargue}},
  \bibinfo {author} {\bibfnamefont {K.}~\bibnamefont {Chance}}, \bibinfo
  {author} {\bibfnamefont {A.}~\bibnamefont {Coustenis}}, \bibinfo {author}
  {\bibfnamefont {B.}~\bibnamefont {Drouin}}, \bibinfo {author} {\bibfnamefont
  {J.}~\bibnamefont {Flaud}}, \bibinfo {author} {\bibfnamefont
  {R.}~\bibnamefont {Gamache}}, \bibinfo {author} {\bibfnamefont
  {J.}~\bibnamefont {Hodges}}, \bibinfo {author} {\bibfnamefont
  {D.}~\bibnamefont {Jacquemart}}, \bibinfo {author} {\bibfnamefont
  {E.}~\bibnamefont {Mlawer}}, \bibinfo {author} {\bibfnamefont
  {A.}~\bibnamefont {Nikitin}}, \bibinfo {author} {\bibfnamefont
  {V.}~\bibnamefont {Perevalov}}, \bibinfo {author} {\bibfnamefont
  {M.}~\bibnamefont {Rotger}}, \bibinfo {author} {\bibfnamefont
  {J.}~\bibnamefont {Tennyson}}, \bibinfo {author} {\bibfnamefont
  {G.}~\bibnamefont {Toon}}, \bibinfo {author} {\bibfnamefont {H.}~\bibnamefont
  {Tran}}, \bibinfo {author} {\bibfnamefont {V.}~\bibnamefont {Tyuterev}},
  \bibinfo {author} {\bibfnamefont {E.}~\bibnamefont {Adkins}}, \bibinfo
  {author} {\bibfnamefont {A.}~\bibnamefont {Baker}}, \bibinfo {author}
  {\bibfnamefont {A.}~\bibnamefont {Barbe}}, \bibinfo {author} {\bibfnamefont
  {E.}~\bibnamefont {Canè}}, \bibinfo {author} {\bibfnamefont
  {A.}~\bibnamefont {Császár}}, \bibinfo {author} {\bibfnamefont
  {A.}~\bibnamefont {Dudaryonok}}, \bibinfo {author} {\bibfnamefont
  {O.}~\bibnamefont {Egorov}}, \bibinfo {author} {\bibfnamefont
  {A.}~\bibnamefont {Fleisher}}, \bibinfo {author} {\bibfnamefont
  {H.}~\bibnamefont {Fleurbaey}}, \bibinfo {author} {\bibfnamefont
  {A.}~\bibnamefont {Foltynowicz}}, \bibinfo {author} {\bibfnamefont
  {T.}~\bibnamefont {Furtenbacher}}, \bibinfo {author} {\bibfnamefont
  {J.}~\bibnamefont {Harrison}}, \bibinfo {author} {\bibfnamefont
  {J.}~\bibnamefont {Hartmann}}, \bibinfo {author} {\bibfnamefont
  {V.}~\bibnamefont {Horneman}}, \bibinfo {author} {\bibfnamefont
  {X.}~\bibnamefont {Huang}}, \bibinfo {author} {\bibfnamefont
  {T.}~\bibnamefont {Karman}}, \bibinfo {author} {\bibfnamefont
  {J.}~\bibnamefont {Karns}}, \bibinfo {author} {\bibfnamefont
  {S.}~\bibnamefont {Kassi}}, \bibinfo {author} {\bibfnamefont
  {I.}~\bibnamefont {Kleiner}}, \bibinfo {author} {\bibfnamefont
  {V.}~\bibnamefont {Kofman}}, \bibinfo {author} {\bibfnamefont
  {F.}~\bibnamefont {Kwabia–Tchana}}, \bibinfo {author} {\bibfnamefont
  {N.}~\bibnamefont {Lavrentieva}}, \bibinfo {author} {\bibfnamefont
  {T.}~\bibnamefont {Lee}}, \bibinfo {author} {\bibfnamefont {D.}~\bibnamefont
  {Long}}, \bibinfo {author} {\bibfnamefont {A.}~\bibnamefont {Lukashevskaya}},
  \bibinfo {author} {\bibfnamefont {O.}~\bibnamefont {Lyulin}}, \bibinfo
  {author} {\bibfnamefont {V.}~\bibnamefont {Makhnev}}, \bibinfo {author}
  {\bibfnamefont {W.}~\bibnamefont {Matt}}, \bibinfo {author} {\bibfnamefont
  {S.}~\bibnamefont {Massie}}, \bibinfo {author} {\bibfnamefont
  {M.}~\bibnamefont {Melosso}}, \bibinfo {author} {\bibfnamefont
  {S.}~\bibnamefont {Mikhailenko}}, \bibinfo {author} {\bibfnamefont
  {D.}~\bibnamefont {Mondelain}}, \bibinfo {author} {\bibfnamefont
  {H.}~\bibnamefont {Müller}}, \bibinfo {author} {\bibfnamefont
  {O.}~\bibnamefont {Naumenko}}, \bibinfo {author} {\bibfnamefont
  {A.}~\bibnamefont {Perrin}}, \bibinfo {author} {\bibfnamefont
  {O.}~\bibnamefont {Polyansky}}, \bibinfo {author} {\bibfnamefont
  {E.}~\bibnamefont {Raddaoui}}, \bibinfo {author} {\bibfnamefont
  {P.}~\bibnamefont {Raston}}, \bibinfo {author} {\bibfnamefont
  {Z.}~\bibnamefont {Reed}}, \bibinfo {author} {\bibfnamefont {M.}~\bibnamefont
  {Rey}}, \bibinfo {author} {\bibfnamefont {C.}~\bibnamefont {Richard}},
  \bibinfo {author} {\bibfnamefont {R.}~\bibnamefont {Tóbiás}}, \bibinfo
  {author} {\bibfnamefont {I.}~\bibnamefont {Sadiek}}, \bibinfo {author}
  {\bibfnamefont {D.}~\bibnamefont {Schwenke}}, \bibinfo {author}
  {\bibfnamefont {E.}~\bibnamefont {Starikova}}, \bibinfo {author}
  {\bibfnamefont {K.}~\bibnamefont {Sung}}, \bibinfo {author} {\bibfnamefont
  {F.}~\bibnamefont {Tamassia}}, \bibinfo {author} {\bibfnamefont
  {S.}~\bibnamefont {Tashkun}}, \bibinfo {author} {\bibfnamefont
  {J.}~\bibnamefont {Vander~Auwera}}, \bibinfo {author} {\bibfnamefont
  {I.}~\bibnamefont {Vasilenko}}, \bibinfo {author} {\bibfnamefont
  {A.}~\bibnamefont {Vigasin}}, \bibinfo {author} {\bibfnamefont
  {G.}~\bibnamefont {Villanueva}}, \bibinfo {author} {\bibfnamefont
  {B.}~\bibnamefont {Vispoel}}, \bibinfo {author} {\bibfnamefont
  {G.}~\bibnamefont {Wagner}}, \bibinfo {author} {\bibfnamefont
  {A.}~\bibnamefont {Yachmenev}},\ and\ \bibinfo {author} {\bibfnamefont
  {S.}~\bibnamefont {Yurchenko}},\ }\bibfield  {title} {\bibinfo {title} {The
  {HITRAN2020} molecular spectroscopic database},\ }\href
  {https://linkinghub.elsevier.com/retrieve/pii/S0022407321004416} {\bibfield
  {journal} {\bibinfo  {journal} {Journal of Quantitative Spectroscopy and
  Radiative Transfer}\ }\textbf {\bibinfo {volume} {277}},\ \bibinfo {pages}
  {107949} (\bibinfo {year} {2022})}\BibitemShut {NoStop}%
\end{thebibliography}

\end{document}